\documentclass[iop]{emulateapj}

\usepackage{natbib}
\usepackage{amsmath}
\usepackage{threeparttable}
\usepackage{amssymb}
\usepackage{graphicx}
\usepackage{longtable}
\usepackage{tabularx}
\newcommand{\HI}{\ensuremath{\mbox{\rm \ion{H}{1}}}}
\newcommand{\HII}{\ensuremath{\mbox{\rm \ion{H}{2}}}}

\newcommand{\htwo}{\ensuremath{\mbox{H$_2$}}}
\newcommand{\msun}{\ensuremath{M_\odot}}

\newcommand{\zsun}{\ensuremath{Z_\odot}}

\newcommand{\sunits}{\mbox{\msun ~pc$^{-2}$}}
\newcommand{\pc}{\ensuremath{\mbox{pc}}}
\newcommand{\kms}{\mbox{km~s$^{-1}$}}
\newcommand{\xco}{\ensuremath{X_{\mathrm{CO}}}}

\newcommand{\ico}{\ensuremath{I_{\mathrm{CO}}}}

\newcommand{\aco}{\ensuremath{\alpha_{\mathrm{CO}}}}
\newcommand{\co}[1]{\mbox{$^{#1}$CO}}

\newcommand{\xunits}{\mbox{cm$^{-2}$ (K km s$^{-1}$)$^{-1}$}}
\newcommand{\aunits}{\mbox{$M_\odot$ (K km s$^{-1}$ pc$^2$)$^{-1}$}}
\newcommand{\counits}{\mbox{K km s$^{-1}$}}

\newcommand{\arc}{\mbox{$^{\prime\prime}$}}

\newcommand{\mlum}{\ensuremath{M_{\rm lum}}}
\newcommand{\mvir}{\ensuremath{M_{\rm vir}}}

\newcommand{\lco}{\ensuremath{L_{\rm CO}}}
\newcommand{\mstar}{\ensuremath{M_\star}}
\newcommand{\mhi}{\ensuremath{M_{\rm HI}}}
\newcommand{\mhtwo}{\ensuremath{M_{{\rm H}_2}}}

\defcitealias{Solomon_1987}{S87}
\defcitealias{Bolatto_2008}{B08}

\shortauthors{Imara et al.}

\begin{document}

\title{ALMA Observations of the Molecular Clouds in NGC 625}

\author{Nia Imara}
\affil{Harvard-Smithsonian Center for Astrophysics, 60 Garden Street, Cambridge, MA 02138}

\author{Ilse De Looze}
\affil{Sterrenkundig Observatorium, Universiteit Gent,
 Krijgslaan 281 S9, B-9000 Gent, Belgium}
\affil{Department of Physics and Astronomy, University College London, Gower Street, London WC1E 6BT, UK}

\author{Christopher M. Faesi}
\affiliation{University of Massachusetts - Amherst, 710 N. Pleasant St., Amherst, MA 01003}
\affiliation{Harvard-Smithsonian Center for Astrophysics, 60 Garden Street, Cambridge, MA 02138}

\author{Diane Cormier}
\affil{AIM, CEA, CNRS, Universit\'e Paris-Saclay, Universit\'e Paris Diderot, Sorbonne Paris Cit\'e, F-91191 Gif-sur-Yvette, France}

\email{nimara@cfa.harvard.edu}

\begin{abstract}
We present the highest resolution (1$\arcsec$) $^{12}$CO observations of molecular gas in the dwarf starburst galaxy NGC 625, to date, obtained with ALMA. Molecular gas is distributed in discrete clouds within an area of $0.4$ kpc$^2$ and does not have well-ordered large-scale motions.  We measure a total molecular mass in NGC 625 of $5.3\times 10^6$ \msun, assuming a Milky Way CO-to-\htwo~conversion factor. We use the CPROPS package to identify molecular clouds and measure their properties.  The 19 resolved CO clouds have a median radius of 20\,pc, a median line width 2.5 \kms, and a median surface density of 169 \sunits. Larson scaling relations suggest that molecular clouds in NGC\,625 are mostly in virial equilibrium.  Comparison of our high-resolution CO observations with a star formation rate map, inferred from ancillary optical observations observations, suggests that about 40\% of the molecular clouds coincide with the brightest H{\sc{ii}} regions.  These bright \HII~regions have a range of molecular gas depletion timescales, all within a factor of $\sim3$ of the global depletion time in NGC\,625 of 106-134\,Myr. The highest surface density molecular clouds towards the southwest of the galaxy, in a region we call the Butterfly, do not show strong star formation activity and suggest a depletion time scale longer than 5\,Gyr. 
\end{abstract}


\section{Introduction}
Nearby star forming dwarf galaxies present a wonderful opportunity for studying the initial conditions of star formation in environments different from the Milky Way.  With their intense radiation fields, low metal abundances, high gas-to-dust ratios, low masses, and shallow potential wells, the conditions in the interstellar medium (ISM) of dwarf galaxies \citep[e.g.,][]{Kunth_2000, Madden_2006, Draine_2007} may affect the properties of giant molecular clouds (GMCs), the sites of most star formation \citep[e.g.,][]{Bolatto_2013}. Moreover, nearby dwarf galaxies are useful laboratories to investigate early Universe star formation which occurred in low-metallicity environments.

A question at the heart of star formation research is how the properties of GMCs---properties including size, mass, velocity dispersion, and surface density---vary across galactic environment \citep[e.g.,][]{Fukui_2010}.  Relatedly, we would like to understand whether and how galactic environment affects the distribution of and correlation between the physical properties of GMCs.  A consensus on the key environmental factors driving differences in cloud character has not been reached.  This is in part due to inhomogeneous data sets used to make comparisons between cloud populations; in part due to our incomplete knowledge of quantities like the CO-to-\htwo~conversion factor, which must be assumed in order to estimate luminous masses for clouds, which are composed primarily of molecular hydrogen, \htwo. In low-metallicity galaxies, the CO luminosity tends to be faint. The intense radiation fields combined with suppressed dust-shielding may lead to the dissociation of molecules at a quicker rate, and the formation of \htwo~on dust grains is expected to be curbed \citep[e.g.,][]{Wolfire_2010}. Moreover, since the CO molecule is more susceptible to dissociating UV radiation compared to \htwo, the CO-emitting regions of molecular clouds in these environments may be smaller and the CO-to-\htwo~conversion factor higher \citep[e.g.,][]{Arimoto_1996, Israel_1997, Boselli_2002, Imara_2007, Leroy_2011, Schruba_2012, Bolatto_2013, Cormier_2014, Hunt_2015, Amorin_2016}.

In recent years, wide-field and high-resolution observations of the molecular ISM have become increasingly available, furthering our knowledge of resolved, extragalactic molecular clouds \citep[e.g.,][]{Bolatto_2008, Fukui_2008, Wong_2011, Leroy_2015, Rubio_2015, Kepley_2016, Schruba_2017, Faesi_2018, Imara_2019}. These studies have demonstrated that although some cloud properties and scaling relations are similar across environments, those in dwarf and starburst galaxies often display considerable differences from their counterparts in massive disk galaxies \citep[e.g.,][]{Leroy_2015, Kepley_2016, Imara_2019}. Our understanding of the contrasts between star-forming gas in starburst dwarf galaxies and disks is particularly limited, as only a handful of these former systems have been observed at sufficiently high resolution and sensitivity to acquire large number statistics of their cloud populations \citep[e.g.,][]{Leroy_2015, Kepley_2016, Imara_2019, Miura_2018}.

Blue compact dwarf (BCD) galaxies---faint, compact objects with high star formation rates relative to their masses---have been proposed as good analogs for starburst systems in the early Universe \citep[e.g.,][]{Leroy_2015, Miura_2018}.  What drives the differences between molecular clouds in these environments and in disks and how star formation is triggered in starbursts are questions under active investigation.  Atacama Large Millimeter/Submillimeter Array (ALMA) observations of the BCD galaxy II Zw 40 revealed clumpy molecular gas characterized by larger line widths and higher cloud surface densities compared to molecular clouds in disk galaxies \citep{Kepley_2016}. More recently, \citet{Imara_2019} observed the BCD Henize 2-10 with ALMA, showing that while the surface densities of GMCs in this galaxy are similar to those of Milky Way clouds, the average molecular gas surface density of the BCD \emph{as a whole} is a factor of 30 to 70 higher than in the Milky Way disk, reflecting the fact that the molecular gas filling factor in Henize 2-10 is close to unity.  The studies of II Zw 40 and Henize 2-10 represent only a small number of existing millimeter observations of BCDs that are available to investigate GMC properties in these extreme environments. 

We present new \co{12}$(J=1-0)$ ALMA observations of the blue compact dwarf NGC 625, a relatively isolated, gas-rich galaxy in the Sculptor Group, at a distance of $3.89\pm 0.22$ Mpc \citep{Cannon_2003}.  NGC 625 has a metallicity of $12+\log(\rm{O/H})=8.22$ \citep[$\approx 1/3\zsun$;][]{Skillman_2003}.\footnote{We assume a solar oxygen abundance of $12+\log(\rm{O/H})=8.69$ \citep{Asplund_2009}.}  The galaxy is one of the few known dwarfs with a large-scale, high-velocity \HI~outflow, which may be a consequence of its prolonged, $\sim 100$ Myr episode of star formation \citep{Cannon_2004, Cannon_2005}.  The stellar mass of NGC 625  \citep[$3\times 10^8\msun$;][]{Madden_2013} and its atomic hydrogen mass \citep[$1.1\times 10^8\msun$;][]{Cannon_2004} are comparable to those of IC 10 (see Table 1).  Single-dish CO observations of NGC\,625 were reported by \cite{Cormier_2014} who found a total molecular gas mass of $5\times10^6$\,\msun, assuming a Galactic CO-to-\htwo~conversion factor. Its current star formation rate (SFR) of $0.05$ \msun~yr$^{-1}$ is similar to that of the SMC, although its gas depletion time of $\sim 3$ Gyr may be twice as long as that in the SMC \citep{Skillman_2003}.  

Our main goals in this study are to present the new data, to describe the global properties of the molecular gas as inferred from the CO observations, and to identify and characterize the properties of molecular clouds in NGC 625. To explore how differences in galactic environment---particularly in starburst dwarf galaxies---influence the properties of molecular clouds, we will examine how empirical relationships determined for NGC 625 molecular clouds compare with general trends observed in the Milky Way and external galaxies. Lastly, we will investigate the relationship between GMCs and star formation activity as inferred from H$\alpha$ observations.  In Section \ref{sec:observations} we provide an overview of our ALMA observations, the highest resolution to date of NGC 625, and we describe the H$\alpha$ observations to which we later compare the distribution of molecular gas.  In Section \ref{sec:global} we describe the global properties of the molecular gas.  In Section \ref{sec:results} we identify and characterize the properties of GMCs in NGC 625, and we compare them to trends observed for molecular clouds in other galaxies.  We discuss the relationship between GMCs and \HII~regions in Section \ref{sec:distribution}, and in \S\ref{sec:sfe} we discuss the star formation efficiency in NGC 625.  We summarize our conclusions in \S\ref{sec:summary}.

\begin{table}\centering
\begin{center}
\begin{tabular}{lcc}
\multicolumn{3}{c}{Table 1: Properties of NGC 625.}\\
\tableline\tableline
Property       & Value      &   Reference \\
\tableline
Distance       & 3.9 Mpc    &  1 \\
Absolute $B$ magnitude  &    $-16.2$     &  2 \\
Metallicity      & $12+\log(\rm{O/H})=8.22$  &  3 \\
Size ($D_{25}$)  & $5\farcm8\times 1\farcm9$  &  4  \\
Stellar mass     & $3.0\times 10^8 \msun$  &  4 \\
FIR luminosity    &   $2.6\times 10^8~L_\odot$  & 4  \\
SFR$_{{\rm FIR}}$    & 0.04 \msun~yr$^{-1}$    &  4 \\
SFR$_{{\rm H}\alpha}$    & 0.05 \msun~yr$^{-1}$    &  3 \\
Inclination      &  $65^\circ$  &  5 \\
\HI~mass       &    $(1.1\pm0.2)\times 10^8 \msun$         &  5 \\ 
\htwo~mass (single-dish)     &    $(5.0\pm0.7)\times 10^6~\msun$        & 6  \\
\htwo~mass (ALMA)     &    $(5.34\pm0.34)\times 10^6~\msun$        & This work  \\
Molecular depletion time   & $106-134$ Myr   &  This work \\
\tableline

\end{tabular}
\caption{(1) \citet{Cannon_2003}; (2) \citet{deVaucouleurs_1991}; (3) \citet{Skillman_2003}; (4) \citet{Madden_2013}; (5) \citet{Cannon_2004}; (6) \citet{Cormier_2014} }
\label{table1}
\end{center}
\end{table}


\section{Observations}\label{sec:observations}
\subsection{New ALMA CO observations}\label{sec:co_observations}
We obtained ALMA Cycle 3 observations (project code 2015.1.01395.S; PI: Nia Imara) in the \co{12}(1-0) line at 115.2712 GHz towards NGC 625 in May 2016.  Two positions were observed that covered the optical nucleus of the galaxy.  The overlapping beams had a phase center of $01^{\rm h}35^{\rm m}06\fs90, -41\degr 26\arcmin13\farcs 47$ [J2000], and the center of the beams was separated by half the half-power beam width ($\sim 25\farcs$).  The ALMA 12-m array was in configuration C36-3  for the five observing nights, with 36 antennas every night, arranged with baselines from 15 m to 639 m, implying a minimum angular resolution of $\sim 1\farcs 1$ and a maximum recoverable scale of $\sim 10\farcs 7$ (at 115.27 GHz). The half-power beamwidth for the 12-m array was $50\farcs7$.  Table 2 summarizes the observing dates, conditions, and calibrators.

\begin{table*}[ht]
\centering
\begin{center}
\begin{tabular}{lccccc}
\multicolumn{6}{c}{Table 2: Observation Summary.}\\
\tableline\tableline
Date                 & 2016 May 3  &  2016 May 7  &  2016 May 7  &  2016 May 8  &  2016 May 8\\
On-source time       & 48.12 min   &  48.15 min  &  48.12 min & 48.13 min  &  48.13 \\
Number of Antennas   & 36        &  36     &  36     &  36     &  36      \\  
Average $T_{\rm sys}$ & 120.6 K  & 128.7 K & 142.6 K & 123.2 K & 160.0 K  \\
Mean precipitable water vapor & 2.73 mm  & 2.90 mm   & 3.07 mm &  2.35 mm & 2.97 mm \\
Bandpass calibrator  & J2357-5311   &  J2357-5311  & J2357-5311 & J2357-5311 & J0334-4088   \\
Flux calibrator      & Neptune   & J2357-5311     & J0334-4008    & J2357-5311 & J0334-4088   \\
Phase calibrator     & J0136-4044   &  J0136-4044  & J0136-4044   & J0136-4044 & J0136-4044  \\
Pointing calibrators & J0134-3843,   &  J2357-5311  & J2357-5311, & J2357-5311 & J0334-4088   \\
                     & J2357-5311, J2246-1206 &  & J0334-4008     &  &  \\
\tableline
\end{tabular}
\label{table2}
\end{center}
\end{table*}


The ALMA Band 3 correlator was set up to have a velocity resolution of 564.453 kHz (1.470 \kms) and a bandwidth of 937.50 MHz, centered at 114.936 GHz to cover the \co{12}(1-0) line, adjusted for the galaxy's LSR velocity of $395$ \kms.  We also observed continuum by centering the remaining basebands at 102 GHz, 101 GHz, and 113 GHz, each with bandwidths of 1.875 GHz.

Our data were processed and imaged using the Common Astronomy Software Applications (CASA) package (https://casa.nrao.edu).  The North American ALMA Science team used CASA version 4.5.3 to manually calibrate the data.  We summarize the data processing steps as follows.  First, there were basic flagging operations, including autocorrelation, shadowed antenna, and edge channel flagging.  Next, a system temperature ($T_{\rm sys}$) calibration table was generated and deviant $T_{\rm sys}$ measurements were flagged.  Then the antenna positions were calibrated, followed by atmospheric calibration using the Water Vapor Radiometer data.  Finally, the bandpass, flux, and gain calibrations were performed.  

We imaged our data in CASA using the \texttt{multiscale} CLEAN algorithm.  This method, which searches for emission at a range of spatial scales, has been shown to do well at recovering extended emission, reducing the depth of negative emission features, and eliminating low-level flux missed by standard CLEAN algorithms, which only clean point source scale emission \citep[e.g.,][]{Rich_2008}.  With the aim of measuring the CO-derived properties of molecular clouds, natural weighting was chosen in order to maximize sensitivity.  We imaged the continuum emission, detected a total flux of about 4 mJy, and subtracted this from the line emission.

Our process for imaging the CO line took several steps: first, we examined each of the 101 (1.5 \kms-wide) velocity channels around the systemic velocity of NGC 625 to determine where significant emission is arising.  Next, a mask was drawn around all the significant and coherent emission.  This mask was then used to deconvolve all image planes. The deconvolution was stopped when the cumulative flux as a function of the number of CLEAN iterations converged.  We confirmed that the deconvolution was satisfactory by confirming that the residual data cube looked like noise.  We performed a third check by repeating the deconvolution process, this time with double the number of CLEAN iterations, and subtracting the resulting deconvolved cubes from the previous results to again confirm that it looked like noise.

Finally, we applied a primary beam correction, using the combined beam pattern of the two pointings, though we note that since most of the emission is concentrated in the central portion of the image, the effect on our measured flux is minimal. The final data cube has voxels with dimensions of $0\farcs 25\times 0\farcs 25 \times 1.5~\kms$ and a spatial extent of $1\farcm 25$. The synthesized beam was measured to be  $1\farcs31\times 1\farcs08$, corresponding to a spatial resolution of 24.6 pc $\times$ 20.3 pc at the distance to NGC 625. The rms sensitivity is $10.3$ mJy per $1.5~\kms$ channel.  The corresponding rms brightness temperature is 0.7 K per channel.  For a typical molecular cloud linewidth of $\sim3$ \kms~(i.e., two channels), this yields an rms integrated intensity of $\sim 2$ \counits.  Assuming a Galactic CO-to-\htwo~conversion factor of 4.35 \aunits~\citep[see \S\ref{sec:intensity} below;][]{Dame_2001, Bolatto_2013}, the corresponding $5\sigma$ sensitivity limit of the mass surface density is $43$ \sunits.

Interferometric observations without zero- or short-spacing data inevitably miss flux, particularly extended emission, from the observed source.  We did not have total power observations, and so we examined the impact of imaging the data with tapered visibilities on flux recovery.  We repeated the deconvolution process using tapered visibility (\emph{uv}) weights, which effectively increases the surface brightness sensitivity at the cost of angular resolution.  There are, however, limits to tapering in the \emph{uv} plane, since the overall sensitivity will decrease as the data are increasingly down-weighted and information at small spatial scales is lost.  We used Gaussian taper functions that yielded final resolutions of roughly $3\arc$ and $6\arc$. The rms noise levels of the  $3\arc$ and $6\arc$ data cubes are 15.0 mJy per $3\arc$ beam and 26.4 mJy per $6\arc$ beam, respectively.

\begin{figure}
 \epsscale{1.1} 
  \plotone{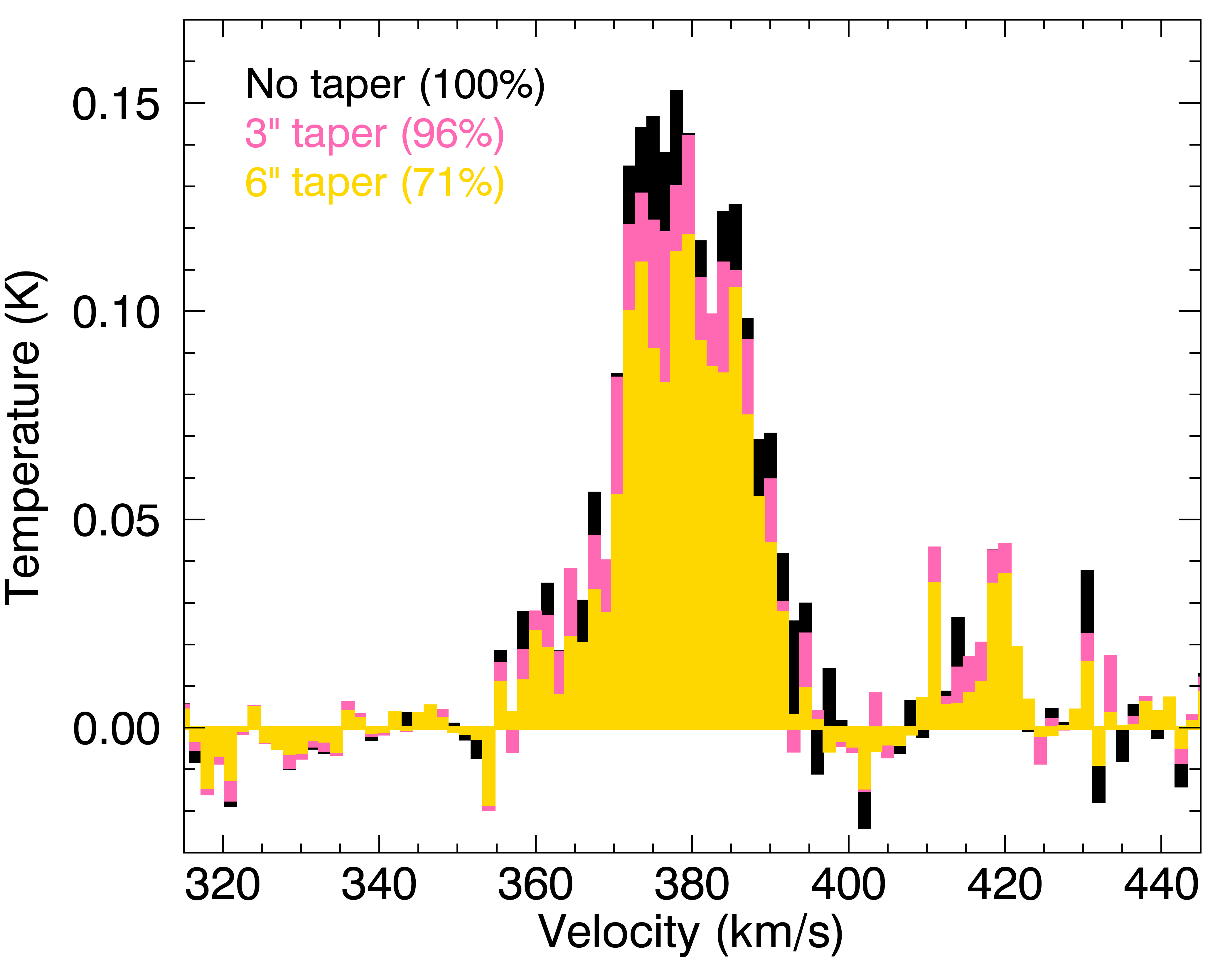}
    \caption{Composite \co{12} spectra of NGC 625.  The black, pink, and gold spectra correspond to data cubes imaged at roughly $1\arc$, $3\arc$, and $6\arc$, respectively.  The top left corner lists the percentage of flux recovered, with respect to the $1\arc$ cube.}
    \label{fig:spectra_compare}
\end{figure}

Figure \ref{fig:spectra_compare} displays the flux recovered by the CLEAN algorithm for the three different data cubes.  The high-resolution, $\sim1\arc$ cube does not miss any significant amount of flux, while the $3\arc$ cube captures 96\% of the total flux, and the $6\arc$ cube captures $71\%$ of the total flux, with respect to the high-resolution cube. Although the tapered cubes recovered flux at extended spatial scales, they suppressed enough information at small scales that the overall recovered flux was less than that in the high-resolution cube.  This suggests that our observations have not suffered significant spatial filtering.  This agrees with our findings, discussed in S\ref{sec:intensity}, that our ALMA observations have a comparable total \co{12} luminosity to previous single dish data obtained with the Australia Telescope National Facility (ATNF) Mopra 22m telescope by \citet{Cormier_2014}.

Figure \ref{fig:optical} in the Appendix shows an optical image of NGC 625 with our observed positions indicated by circles.  Figure \ref{fig:peak_flux}, also in the Appendix, displays the ALMA CO(1-0) peak flux map.

\subsection{Archival Hubble H$\alpha$ observations}\label{sec:hubble}
In Sections \ref{sec:distribution} and \ref{sec:sfe} we compare the distribution of molecular gas to H$\alpha$ observations probing recent star formation activity.  NGC\,625 was observed with the Wide Field Planetary Camera 2 (WFPC2) on the \emph{Hubble Space Telescope} (HST) by \citet{Cannon_2003}.  We retrieved images from the \textit{Hubble} Legacy Archive in four passbands, including narrow-band images H$\alpha$ (F656N filter) and H$\beta$ (F487N filter), and broad-band images in the visible (F555W filter) and near-infrared (F814W filter). These \textit{Hubble} images were (manually) corrected for astrometry offsets (with a final accuracy of $\sim 0\farcs1$ for the astrometry) based on a comparison with the astrometry of stars in the NOMAD catalog, and then flux calibrated based on the PHOTFLAM keyword in the respective headers of the images.

To obtain continuum-subtracted H$\alpha$ and H$\beta$ maps, we first subtracted the combined H$\alpha$ and H$\beta$ emission (with a contribution of only a few percent) from the F555W continuum map, until no diffuse extended emission was visible in the F555W map and the morphology was similar to the line-free F814W continuum map, following the strategy suggested by \citet{Cannon_2002, Cannon_2003}. The line-free F555W map was subsequently used to obtain continuum-subtracted narrow-band H$\alpha$ and H$\beta$ line maps. 

The images were corrected for Galactic extinction following the reddening $E(B-V)=0.16$\,mag reported by \citet{Schlegel_1998}, assuming a Galactic dust extinction law. We applied the Balmer decrement (i.e., the ratio of the H$\alpha$/H$\beta$ lines) to correct for internal dust extinction, following \citet{Kreckel_2013}. Due to the low signal-to-noise of the H$\beta$ line, the internal dust extinction correction was mostly restricted to the two brightest H{\sc{ii}} regions \citep[called A and B in the nomenclature of][]{Cannon_2003} with high H$\alpha$ and H$\beta$ equivalent width \citep{Cannon_2003}. Based on the peak of MIPS\,24\,$\mu$m emission in those two brightest H{\sc{ii}} regions (see Figure \ref{fig:halpha_spitzer} in the Appendix), and the weak mid-infrared emission originating from other parts of NGC\,625, we argue that the dust extinction correction can be considered negligible outside of these two H{\sc{ii}} regions.

Theoretically, the (unobscured) H$\alpha$/H$\beta$ ratio is equal to 2.86, under the assumption of Case B recombination, a gas temperature of T=10,000\,K, and electron density $n_{\text{e}}$=100\,cm$^{-3}$ \citep{Osterbrock_1989}. On average, we find H$\alpha$/H$\beta$ ratios of 3.7 and 3.6 in the brightest H{\sc{ii}} regions A and B, but there is quite some variation throughout both regions. From these H$\alpha$/H$\beta$ ratios, we infer an average $V$-band dust extinction of $A_{\text{V}}$=0.7 and 0.6 for H{\sc{ii}} regions A and B, respectively. Our measured dust extinction is consistent with the $A_{\text{V}}$ values of around 0.5 quoted in \citet{Cannon_2003} for these two regions. The total SFR (=0.5\,M$_{\odot}$ yr$^{-1}$) measured from the H$\alpha$ map after correcting for dust extinction is furthermore in excellent agreement with the SFR estimate from \citet{Cannon_2003}.  

We did not correct for possible N{\sc{ii}} contamination in the H{\sc{ii}} emission.  In the star-forming regions of disk galaxies, at least, there is evidence that the N{\sc{ii}} flux is negligible \citep[e.g.,][]{James_2005}. In particular for NGC\,625, \citet{Skillman_2003b} showed that the nitrogen abundance is very low, and that an uncertainty of about 6$\%$ is introduced to the H$\alpha$ fluxes by not correcting them for possible N{\sc{ii}} contamination.  Nevertheless, if N{\sc{ii}} emission is present in non-negligible amounts in the regions we investigate in NGC 625, we may slightly overestimate star formation rates.

\section{Global Properties of Molecular Gas}\label{sec:global}
\subsection{Integrated intensity}\label{sec:intensity}

\begin{figure}
  \plotone{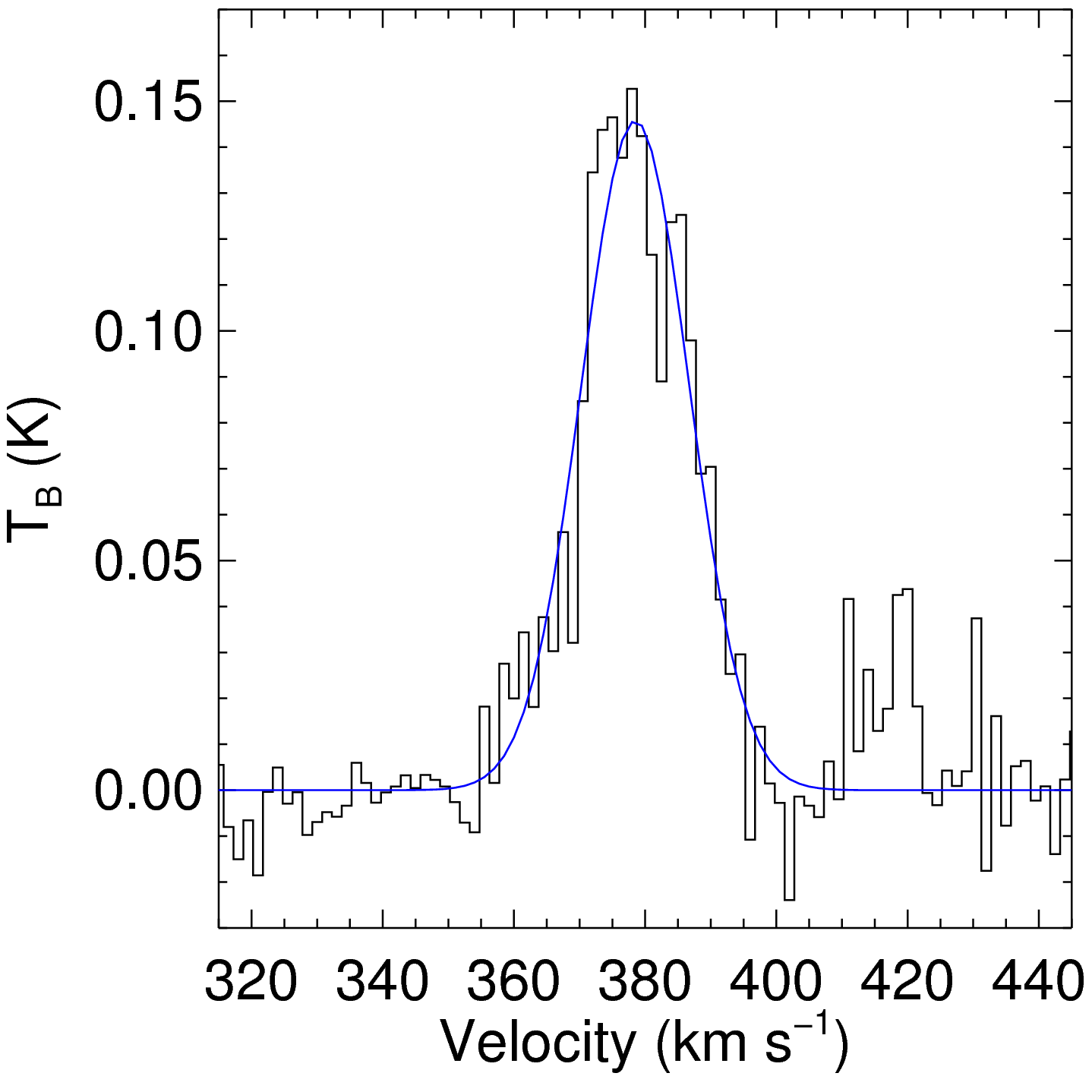}
    \caption{Composite \co{12} spectrum of NGC 625 (black) with a Gaussian fit (blue).}
    \label{fig:spectrum}
\end{figure}

\begin{figure*}[ht]
    \centering
    \includegraphics[width=6.25in]{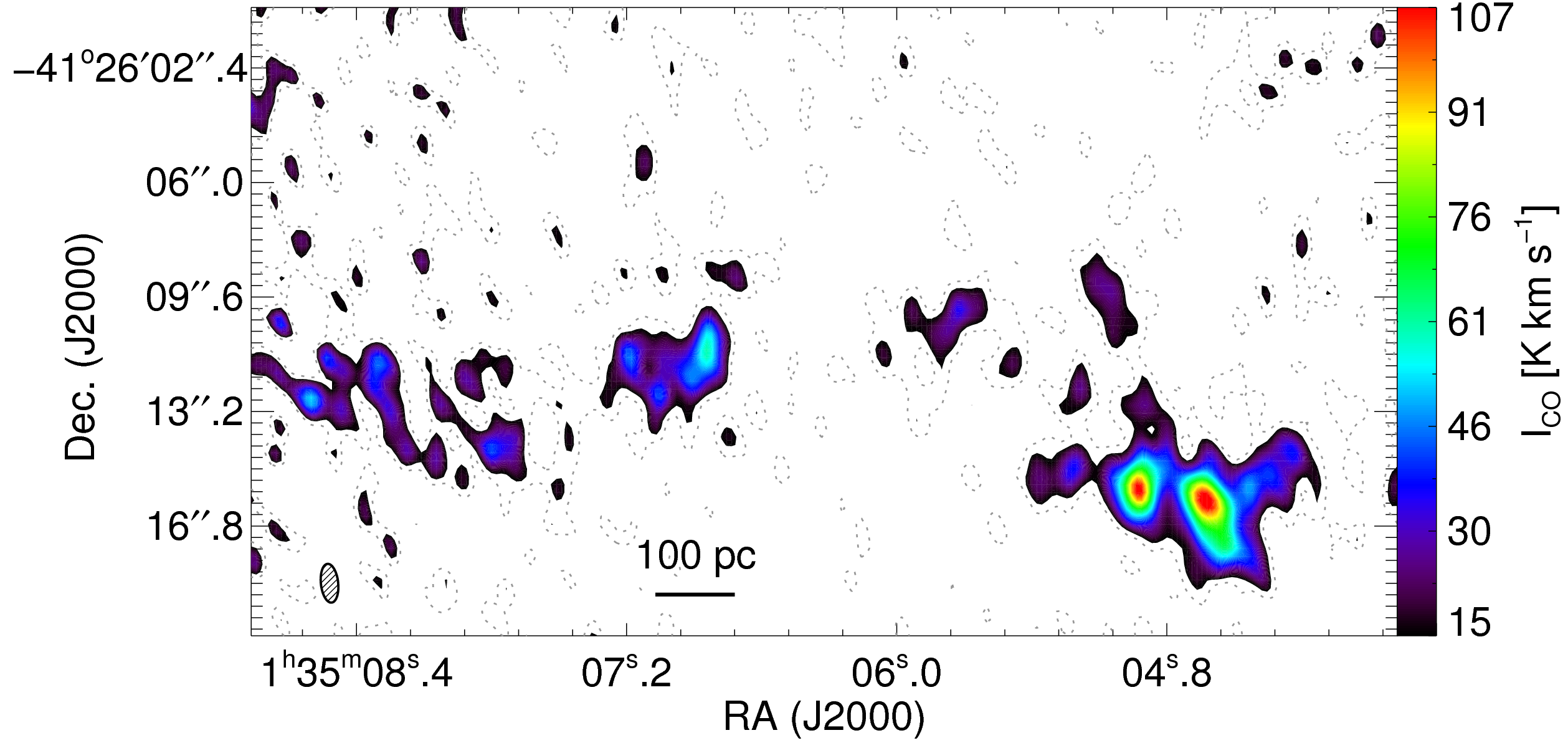}
    \caption{Total integrated intensity (zeroth moment) ALMA \co{12}(1-0) map of NGC 625, integrated over the velocity range 350 to 435 \kms.  The dotted contours represent the $1\sigma_{\rm rms}$ noise level of the map, where $\sigma_{\rm rms}=8.5$ \counits.  The range of the color scale is $2$-$14\sigma_{\rm rms}$.  The $1\farcs08\times 1\farcs31$ synthesized beam is indicated in the lower left.}
    \label{fig:map}
\end{figure*}

In Figure \ref{fig:spectrum} we display the observed integrated \co{12}(1-0) spectrum of NGC 625 in units of brightness temperature, $T_B$, created by averaging $T_B$ across the image of the galaxy (displayed in Figure \ref{fig:map}) in each channel.  We fit a Gaussian to the spectrum, and we measure a peak temperature of $146\pm6$ mK occurring at a systematic central velocity of $378.5$ \kms.  The fit gives a gas velocity dispersion of $(8.20\pm 0.38)$ \kms, yielding a total integrated CO intensity of $I_{\rm CO}=(3.00\pm 0.19)$ \counits.  

\cite{Cormier_2014} reported single-dish CO(1-0) observations obtained with the ATNF Mopra 22-m telescope over the central part of NGC\,625. They covered the galaxy with 4 overlapping pointings with a beamsize of 30\arcsec~corresponding to an area of $\simeq0.9$\,kpc$^2$. They found a luminosity of $(1.1\pm0.2)\times10^6$\,\counits\,pc$^2$, and a molecular gas mass of $(5.0\pm0.7)\times10^6$\,\msun, assuming a Galactic conversion factor.  While the Mopra observations are not as sensitive as the ALMA observations, \citet{Cormier_2014} made measurements on the stacked spectra of the 4 positions observed with Mopra.  These 4 positions roughly correspond to the area over which we measure the average CO spectrum with our ALMA data.  The values obtained by \citet{Cormier_2014} are in good agreement with our total values (see also \S\ref{sec:intensity}), suggesting that our interferometric data have not suffered from significant spatial filtering.

Figure \ref{fig:map} displays the \co{12} integrated intensity map, integrated over the velocity range 350 to 435 \kms.  The size of the region is $57\farcs5\times20\arc$, corresponding to an area of 0.4 kpc$^2$ at the distance to the galaxy.  Given the total integrated CO intensity we measure, the luminosity of this region is $1.2\times10^6$ \counits\,pc$^2$. This is consistent with the luminosity measured by \citet{Cormier_2014}, implying that all the molecular gas in NGC 625 is concentrated in the nucleus of the galaxy and that our data do not miss significant extended emission.

The morphology of the molecular gas is elongated and clumpy, with the two brightest emission peaks located in the western part of the galaxy. Converting from CO luminosity, \lco, to luminous mass, \mlum, requires a CO-to-\htwo~conversion factor,
\begin{equation}
    \alpha_{\rm CO}\equiv\mlum/\lco.
\end{equation}
Equivalently, $\xco\equiv N(\htwo)/\ico$, where $N(\htwo)$ is the \htwo~column density.  The conversion factor typically used for the Milky Way is $\alpha_{\rm CO}=4.35$ \aunits~\citep{Dame_2001, Bolatto_2013}.  Assuming a Galactic conversion for NGC 625, the total luminosity we measure corresponds to a total molecular mass of $(5.34\pm0.34)\times 10^6$ \msun.

\subsection{Dynamics}\label{sec:dynamics}

\begin{figure*}
    \centering
    \includegraphics[width=6.25in]{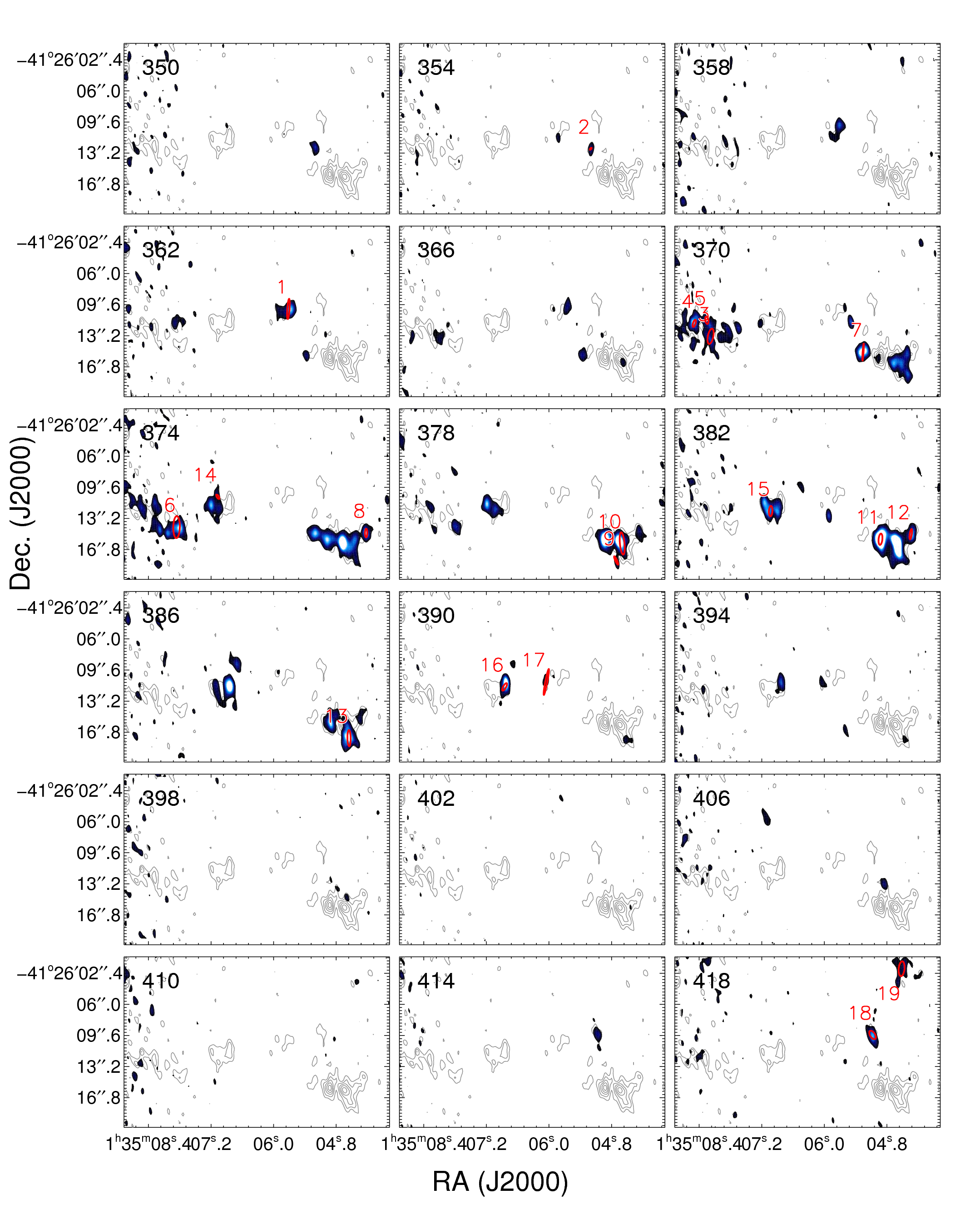}
    \caption{ALMA \co{12} channel maps with velocity widths of 4 \kms, shown in color.  The color scale goes from $0.5\sigma_{\rm rms}$ to $2.25\sigma_{\rm rms}$, where $\sigma_{\rm rms}=8.5$ \counits~is the rms noise level of the intensity map of the entire galaxy, integrated over the full velocity range.  For context, the contour map shows the structure of the entire galaxy integrated over all velocities.}
    \label{fig:channel}
\end{figure*}

In Figure \ref{fig:channel} we display 18 velocity channel maps, each $4$ \kms-wide, from 360 to 395 \kms.  Contour levels are multiples of the rms noise of the zeroth-moment intensity map ($\sigma_{\rm rms}=7.61$ \counits) displayed in Figure \ref{fig:map}.  In each channel, the contours start at $0.25\sigma_{\rm rms}$.  Overplotted are the positions and position angles of GMCs identified in \S\ref{sec:properties}. The bulk of the molecular gas emits strongly between about 366 to 390 \kms.

A systematic velocity gradient across the galaxy is not apparent in the channel maps or in the first-moment map.  We also investigated position velocity diagrams at a number of different cuts through the data set and found no evidence for large-scale rotation or other systematic motions (e.g., outflows) of the molecular gas in NGC 625.  This is interesting because \citet{Cannon_2004} find observational evidence that the complex velocity structure of the \HI~(at a resolution of $\sim6\arc$) is a signature of a large-scale outflow, overlapping with a disk undergoing solid-body rotation.  If, as \citet{Cannon_2004} suggest, the \HI~blowout is due to a prolonged, widespread star formation episode that has delivered sufficient energy into the ISM to drive \HI~into the galactic halo, it seems reasonable to expect that the molecular gas would also be similarly influenced. However, we only detect CO in the disk of the galaxy and do not see clear signatures of the blowout in the CO properties.

\section{GMCs in NGC 625}\label{sec:results}
In the following sections, we generate a catalog of molecular clouds in NGC 625, calculate their properties, and examine empirical relationships from their derived properties.

\subsection{GMC Identification}\label{sec:identification}
We used the \texttt{CPROPS} algorithm \citep{Rosolowsky_2006} to find clouds in our $1\arc$ data cube and measure their properties.  \texttt{CPROPS} has been described in detail in \citet{Rosolowsky_2006} and has been discussed in several recent studies that used the package \citep[e.g.,][]{Colombo_2014, Kepley_2016, Faesi_2018}.  In the following, we briefly summarize the algorithm and the parameters we chose to create the NGC 625 cloud catalog.

\texttt{CPROPS} begins by defining regions of significant emission within a three-dimensional data cube, by identifying pairs of adjacent pixels with signal-to-noise above some threshold level $t\sigma$, where $\sigma$ is the rms noise level per channel.  These regions are expanded to include all adjacent pixels down to a lower threshold level $e\sigma$.    We adopted the default values, $t=4$ and $e=2$, similar to previous studies.  Next, the algorithm identifies ``islands" of emission within these regions that have an area of at least one telescope beam and that include at least two consecutive velocity channels.  Finally, structures resembling molecular clouds are identified from within these islands.  CPROPS searches for local emission peaks by scanning the islands with a moving box having dimensions $15~\pc\times 15~\pc\times 2~\kms$.  Individual clouds are picked out as objects that are above at least $2\sigma$ the shared contour with any neighboring maxima.  In addition to the standard, default parameters, we used the modified \texttt{CLUMPFIND} \citep{Williams_1994} parameter, which requires that all detected emission is assigned to a detected cloud.

To minimize observational bias due to limited sensitivity, CPROPS extrapolates the measurements of cloud properties down to what is expected for finite sensitivity.  This results in extrapolated spatial moments along the major and minor axes of the cloud, $\sigma_{maj,min}(0\rm{K})$, and along the velocity axis, $\sigma_v(0\rm{K})$.    Additionally, CPROPS makes corrections to account for finite angular and velocity resolution of the data, by deconvolving the beam and channel width from measured cloud sizes and line widths.  The algorithm estimates uncertainties in cloud properties with the bootstrapping method.  We used 50 bootstrap iterations to estimate the uncertainties.  The extrapolation, deconvolution, and bootstrapping methods are fully explained in \cite{Rosolowsky_2006}.  For further detail on how using different parameter choices in CPROPS can affect cloud properties in the final catalog, also see \citet{Faesi_2018}.

\begin{figure}
    \epsscale{1.1}
    \plotone{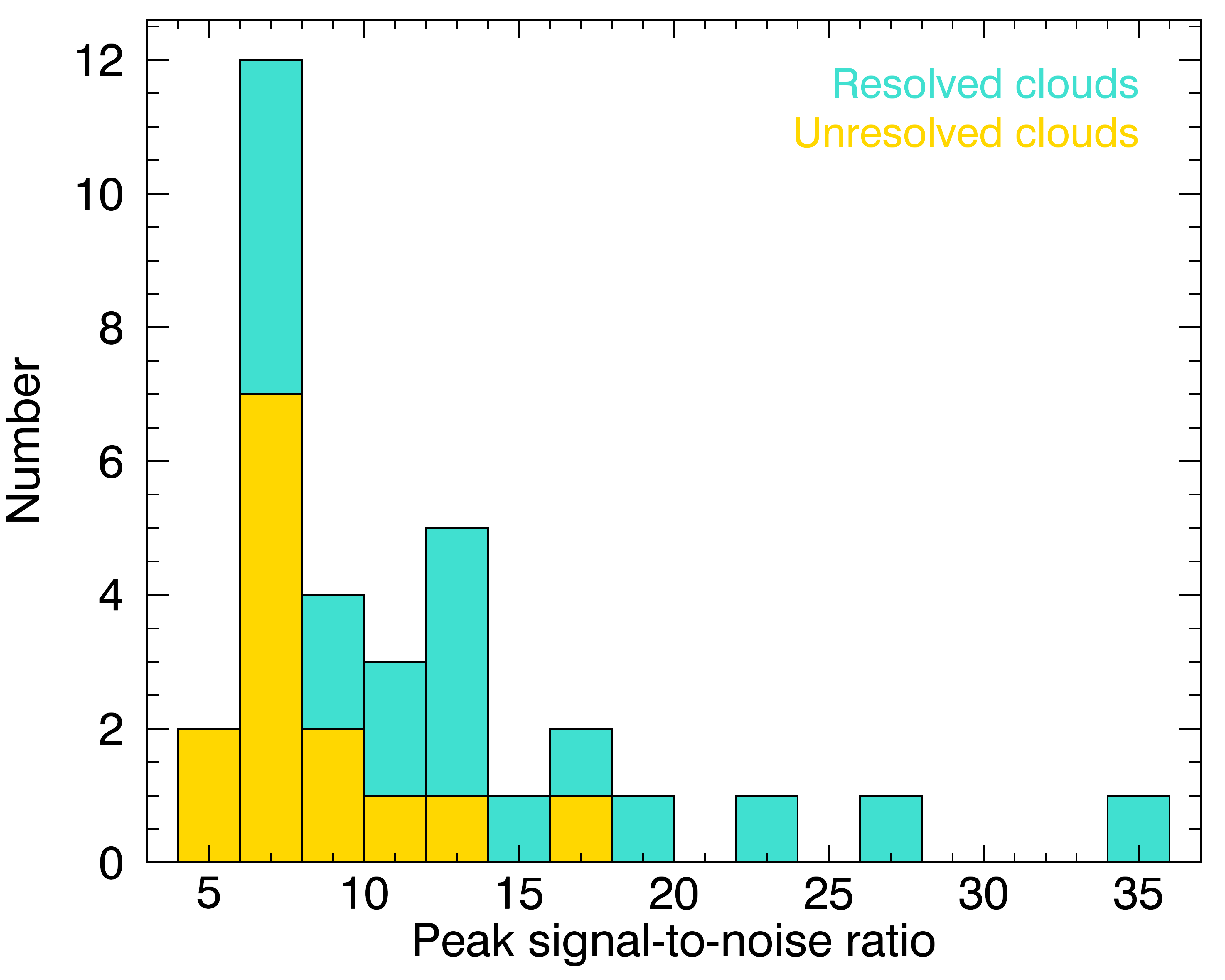}
    \caption{Histogram of the peak signal-to-noise ratio of GMCs.  The turquoise and yellow bars represent resolved and unresolved clouds, respectively.}
    \label{fig:snr}
\end{figure}

\begin{figure*}
    \centering
    \includegraphics[width=6.25in]{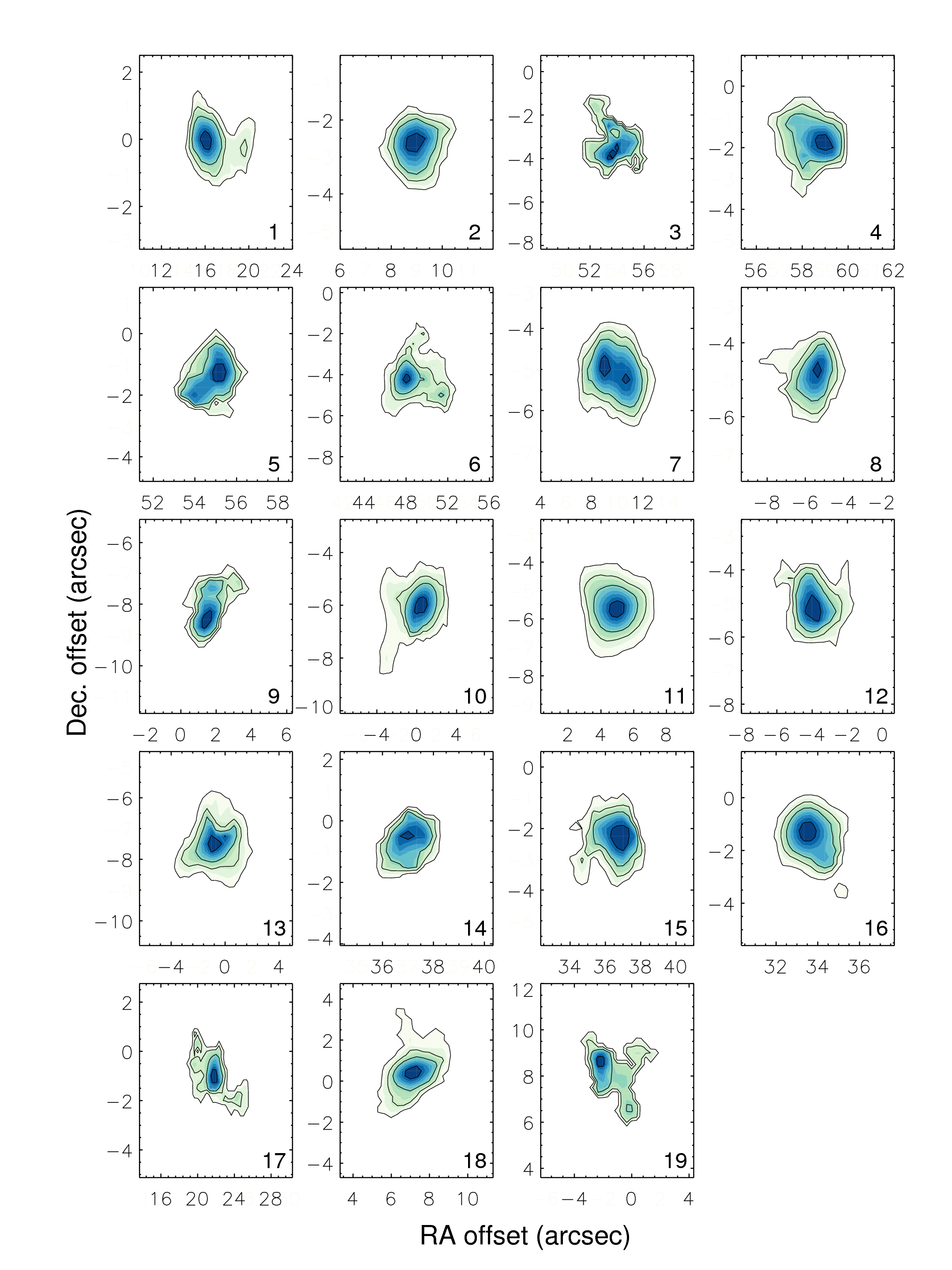}
    \caption{Integrated intensity maps of individual GMCs in NGC 625.} \label{fig:multi_plot}
\end{figure*}

\begin{figure*}
    \centering
    \includegraphics[width=6.25in]{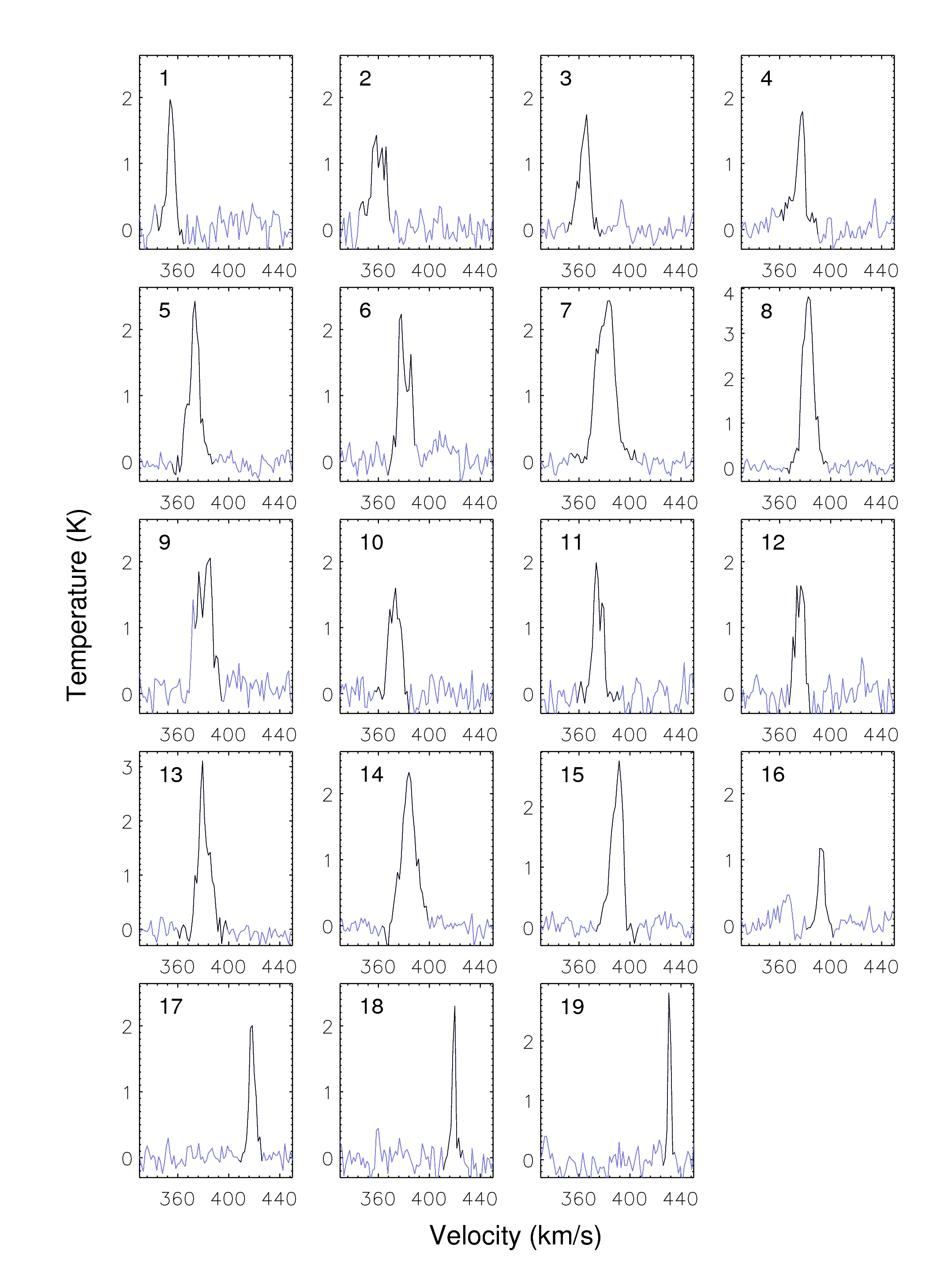}
    \caption{Spectra of individual GMCs in NGC 625.} \label{fig:multi_spec}
\end{figure*}

\begin{table*}[ht]
\centering
\begin{center}
\begin{tabular}{ccccccccc}
\multicolumn{9}{c}{Table 3: Properties of NGC 625 GMCs.}\\
\tableline\tableline
Cloud & R.A. & Dec. & $v_0$ & $R$ & $\sigma_v$ & $M_{\rm lum}$ & $M_{\rm vir}$ & $T_{\rm max}$ \\
      & \multicolumn{2}{c}{(J2000)} & (\kms)& (pc)& (\kms) & ($10^5\msun$) & ($10^5\msun$) & (K) \\
\hline
1  &  01:35:05.72 & -41:26:10.1  &  364.6 &  $22.2\pm2.7$ &  $3.2\pm0.5$ &  $2.5\pm0.1$ &  $2.4\pm0.7$  & 4.5 \\
2  &  01:35:05.20 & -41:26:12.7  &  354.5 &  $ 4.3\pm6.3$ &  $2.6\pm0.5$ &  $0.8\pm0.1$ &  $0.3\pm0.6$  & 3.8  \\
3  &  01:35:08.18 & -41:26:13.3  &  372.2 &  $30.2\pm7.6$ &  $3.0\pm0.9$ &  $1.7\pm0.5$ &  $2.9\pm1.9$  & 3.5 \\
4  &  01:35:08.49 & -41:26:11.8  &  373.7 &  $14.3\pm5.1$ &  $2.1\pm0.8$ &  $1.4\pm0.3$ &  $0.7\pm0.5$  & 4.9 \\
5  &  01:35:08.25 & -41:26:11.4  &  373.5 &  $15.9\pm6.0$ &  $1.8\pm0.7$ &  $0.9\pm0.3$ &  $0.5\pm0.5$  & 3.7 \\
6  &  01:35:07.86 & -41:26:14.2  &  376.2 &  $39.6\pm4.3$ &  $2.8\pm0.3$ &  $3.5\pm0.3$ &  $3.2\pm0.9$  & 6.4 \\
7  &  01:35:05.26 & -41:26:15.1  &  372.8 &  $12.8\pm1.9$ &  $4.1\pm0.4$ &  $3.1\pm0.3$ &  $2.2\pm0.6$  & 6.7 \\
8  &  01:35:04.23 & -41:26:14.9  &  377.9 &  $14.7\pm7.0$ &  $2.1\pm0.5$ &  $1.1\pm0.2$ &  $0.7\pm0.5$  & 5.5 \\
9  &  01:35:04.71 & -41:26:18.1  &  381.2 &  $13.2\pm4.6$ &  $2.9\pm0.8$ &  $0.8\pm0.2$ &  $1.2\pm0.8$  & 2.9 \\
10 &   01:35:04.60 & -41:26:16.1 &   378.4 &  $33.4\pm3.1$ &  $5.1\pm0.3$ &  $10.3\pm0.6$ &  $8.9\pm1.3$ &  9.9 \\
11 &   01:35:04.92 & -41:26:15.6 &   382.5 &  $21.4\pm2.8$ &  $4.0\pm0.3$ &  $7.4\pm0.5$  & $3.6\pm0.8$  & 11.8 \\
12 &   01:35:04.34 & -41:26:15.0 &   384.6 &  $15.3\pm6.0$ &  $2.3\pm0.4$ &  $1.4\pm0.1$  & $0.8\pm0.6$  &  4.3 \\
13 &   01:35:04.55 & -41:26:17.5 &   387.0 &  $23.8\pm3.4$ &  $3.8\pm0.5$ &  $3.9\pm0.3$  & $3.6\pm1.0$  &  5.9 \\
14 &   01:35:07.07 & -41:26:10.7 &   374.7 &  $10.4\pm7.6$ &  $1.5\pm0.8$ &  $0.4\pm0.3$  & $0.3\pm0.3$  &  2.4 \\
15 &   01:35:07.03 & -41:26:12.3 &   382.5 &  $19.8\pm3.6$ &  $4.7\pm0.7$ &  $2.7\pm0.3$  & $4.6\pm1.7$  &  4.9 \\
16 &   01:35:06.85 & -41:26:11.5 &   390.0 &  $16.3\pm3.4$ &  $3.8\pm0.5$ &  $3.7\pm0.2$  & $2.4\pm0.9$  &  8.3 \\
17 &   01:35:06.05 & -41:26:11.0 &   392.8 &  $24.4\pm5.1$ &  $2.1\pm0.4$ &  $1.1\pm0.2$  & $1.1\pm0.7$  &  2.9 \\
18 &   01:35:05.08 & -41:26:09.5 &   418.5 &  $22.8\pm4.9$ &  $2.3\pm0.3$ &  $1.7\pm0.2$  & $1.2\pm0.4$  &  5.3 \\
19 &   01:35:04.52 & -41:26:01.9 &   419.3 &  $30.6\pm8.6$ &  $1.3\pm0.4$ &  $0.8\pm0.2$  & $0.5\pm0.5$  &  2.9 \\
\tableline
\end{tabular}
\label{table3}
\end{center}
\end{table*}


\subsection{GMC Properties}\label{sec:properties}
Using the \texttt{CPROPS} parameters discussed above, we catalog 33 clouds.  Fourteen clouds have deconvolved radii smaller than the telescope beam and by this definition are unresolved \citep{Rosolowsky_2006}.  The remaining 19 resolved clouds make up our final catalog.  Their properties are summarized in Table 3.  Figure \ref{fig:snr} displays a histogram of the peak signal-to-noise ratio (SNR) of all clouds, both resolved and unresolved.  Figure \ref{fig:multi_plot} displays the intensity maps of individual clouds identified by  \texttt{CPROPS}.  And Figure \ref{fig:multi_spec} shows the average \co{12} spectrum along the line of sight toward each cloud.

CPROPS defines the effective radius $R$ of a cloud by subtracting the rms beam size, $\sigma_{beam}$, from the extrapolated spatial moments in quadrature:
\begin{equation}\label{eq:radius}
    R=\sqrt{[\sigma_{maj}^2(0{\rm K})-\sigma_{beam}^2]^{1/2} [\sigma_{min}^2(0{\rm K})-\sigma_{beam}^2  ]^{1/2}}.
\end{equation}
The one-dimensional velocity dispersion, $\sigma_v$, of a cloud is determined by deconvolving the channel width $\Delta V_{chan}$, from the extrapolated second moment:
\begin{equation}\label{eq:sigma_v}
    \sigma_v=\sqrt{\sigma_v^2(0{\rm K})-\frac{\Delta V_{chan}^2}{2\pi}}.
\end{equation}
We will refer to the one-dimensional velocity dispersion and the linewidth ($\sigma_v = \Delta V_{\rm FWHM}/2.355$) interchangeably throughout the paper.

The CO luminosity \lco, virial mass, luminous mass,  and surface density are calculated as follows:
\begin{equation}\label{eq:mvir}
    \mvir = 1040 R\sigma_v^2,
\end{equation}
\begin{equation}\label{eq:mlum}
    \mlum = \alpha_{\rm CO} L_{\rm CO},
\end{equation}
\begin{equation}\label{eq:surf}
    \Sigma = \frac{\mlum}{\pi R^2},
\end{equation}
where $R$ and $\sigma_v$ are in units of parsecs and \kms, and assuming the accepted Milky Way value for the CO-to-\htwo~conversion factor, \aco.  Since our knowledge of the actual CO-to-\htwo~conversion factor in NGC 625 is incomplete (see \S\ref{sec:virial} and \S\ref{sec:global}), we emphasize that the luminous masses we derive correspond to the CO-emitting regions of GMCs.  The coefficient in Equation (\ref{eq:mlum}) for \mvir~assumes that clouds are spherically symmetric and have power law volume density ($\rho$) profile $\rho \propto R^{-1}$ \citep{Solomon_1987,Bolatto_2008}. 

The median values for the size, line width, luminous mass, and surface density of the GMCs are $20$ pc, $2.5$ \kms, $1.7\times 10^5$ \msun, and 169 \sunits.  The properties of all clouds are listed in Table 3.

Following \citet{Faesi_2018}, who in their study of NGC 300 molecular clouds found that GMC sizes smaller than half the beamsize were not well recovered, we adopt a resolution limit of $\sim 11$ pc for our data. We also perform a simple test to verify that the GMCs we identify are not noise peaks. This procedure precisely follows that outlined in \cite{Utomo_2015} (Appendix C; see also \cite{Engargiola_2003}). We treat the probability of a false detection to be given by Poisson statistics, i.e. $P_n(k) = [0.5 \times \rm{erfc}(k/\sqrt{2}]^n$, where $n$ is the number of adjacent channels in a cloud with signal greater than $k\sigma$, where $\sigma$ is the RMS noise. The probability of a real detection is then given by $1 - N_{\rm trial} P_n(k)$ for $N_{\rm trial}P_n(k) \ll 1$, where $N_{\rm trial} = N_{\rm pix}/n$ and $N_{\rm pix} \approx 7.7\times10^4$ is the number of pixels in the cube. Since the \texttt{EDGE} parameter in \texttt{CPROPS} is set to $2$, $k=2$ in the above. For $n=4$, the probability of a real detection is thus 0.92, and for $n=5$ it is 0.996. The resolved cloud with the fewest channels has $n=5$, thus it is very unlikely any of our clouds are misidentified noise peaks.

\begin{figure}
    \epsscale{1.1}
    \plotone{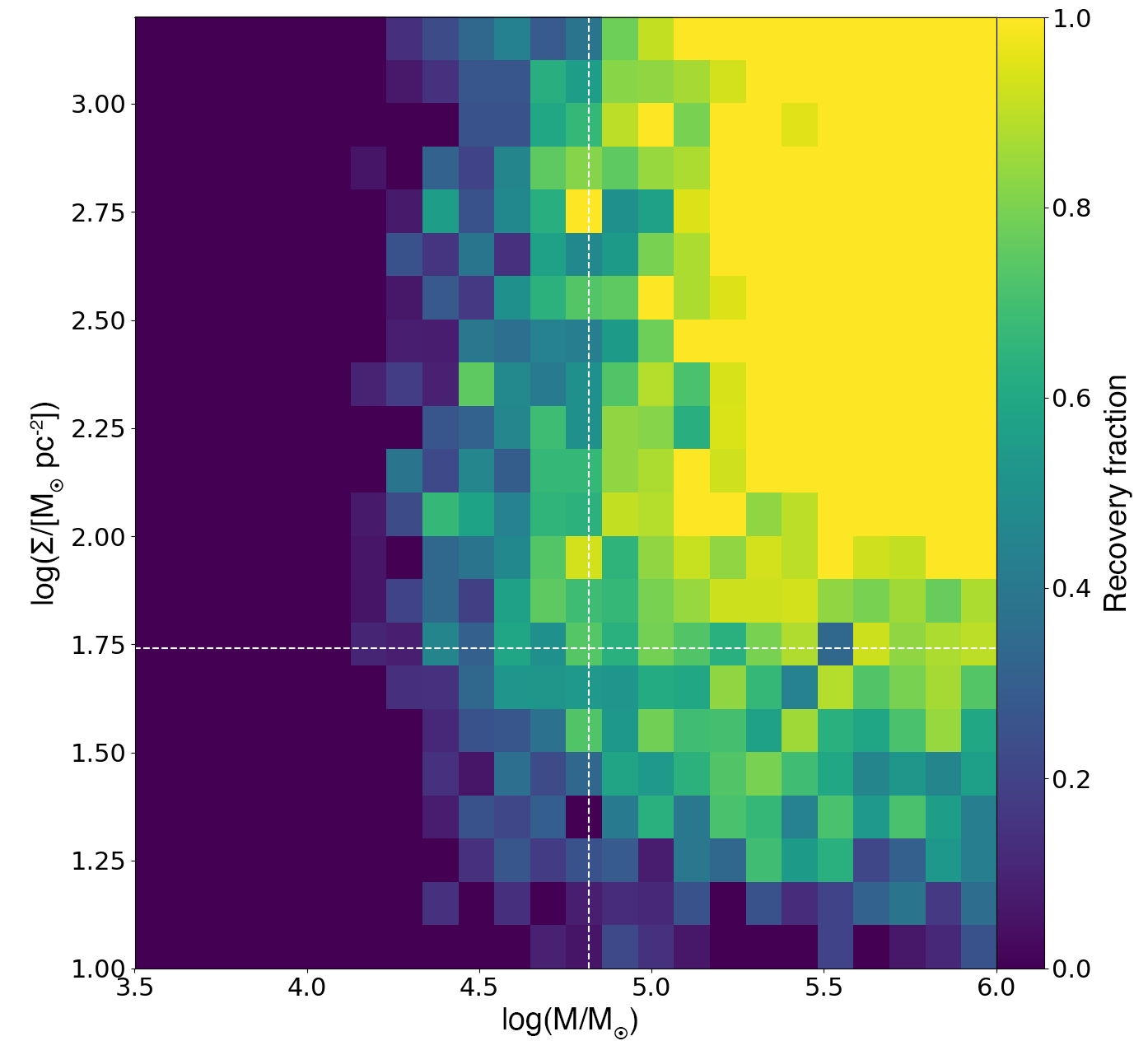}
    \caption{Illustration of the results of our completeness text. The colors cale shows the recovery fraction of artificial sources as a function of mass and surface density. White dashed lines show our adopted completeness limits of $6.5\times10^4~\msun$ and $55~\msun$~pc$^{-2}$.} \label{fig:completeness}
\end{figure}

To more robustly estimate the completeness of our cloud catalog incorporating the effects of the decomposition algorithm, we have conducted a series of false source injection and recovery tests. For each iteration, we generated a set of 6 to 10 molecular ``clouds" with masses in the range [$10^{3.5}$, $10^6$] \msun, surface densities in the range [$10$, $2000$] \msun pc$^{-2}$, and virial parameters in the range $-1 < \log(\alpha_{\rm vir}) < 1$. Each parameter was independently and randomly chosen from a logarithmic distribution spanning the above ranges. The mass was converted to flux using a Milky Way CO-to-\htwo~conversion factor and the distance of NGC~625, and then the flux was distributed as a three-dimensional Gaussian in position-position-velocity space according to the cloud's properties. These false sources were then placed randomly within a data cube with noise properties and dimension equivalent to those of our actual observations. We then ran CPROPS on this false source data cube and compared the recovered cloud positions with the false source positions. If the recovered cloud center fell within the footprint of the injected source, we considered this a successful recovery. We conducted 1000 iterations of this process, resulting in approximately 7000 false sources injected in total. We then calculate the recovery fraction as the number of correctly recovered clouds as a function of their parameters. Figure~\ref{fig:completeness} shows a histogram of the recovery fraction as a function of surface density and mass, the two parameters that appear to have the largest joint impact on source recovery. We marginalize over each parameter individually to define our fiducial 50\% completeness limits of $6.5\times10^4~\msun$ and $55~\msun$~pc$^{-2}$, noting that BOTH these quantities must be greater than these limits in order for a source to be reliably recovered in our tests.

\subsection{The Larson Relations in NGC 625}\label{sec:larson}
We now investigate the scaling relations between cloud properties frequently called ``Larson's Laws" \citet{Larson_1981}.   Millimeter observations show that Milky Way GMCs obey strong trends between size, velocity dispersion, luminosity, and virial mass \citep[e.g.,][]{Larson_1981, Solomon_1987}.  Observed trends for Milky Way clouds have often been used as a benchmark of comparison for extragalactic GMC populations.  The first law, often referred to as the size-linewidth relation, says that the velocity dispersion of molecular clouds increases with size, according to $\sigma_v\sim R^{0.5}$, for Galactic clouds.  The common interpretation of this relation is that the internal turbulence of clouds increases with cloud size.  The second law states that the cloud luminous mass has a nearly one-to-one relation with virial mass, implying that clouds are in virial equilibrium.  The third law says that cloud mass scales with size, such that clouds have roughly constant surface density.  

In the next sections, we examine scaling relations for NGC 625 molecular clouds and  compare them to the empirical relations determined by \citet[][hereafter S87]{Solomon_1987} for GMCs in the inner Milky Way and first described by \cite{Larson_1981}.  We also consider trends derived for extragalactic molecular clouds  by \citet[][hereafter B08]{Bolatto_2008}.  For each set of scaling relations in NGC 625, we calculate the Spearman's rank correlation coefficient, $r_s$, to estimate the degree of correlation between the cloud properties.  If none of the data values are repeated, a coefficient of $r_s=1$ signifies a perfectly monotonically increasing function, while values close to zero indicate little or no correlation.  We fit correlations using the ``BCES" (bivariate, correlated errors with intrinsic scatter) method of \citet{Akritas_1996}.  On each of the following plots displaying a Larson relation, we show the Spearman's coefficient as well as the derived fit, whether or not there is a significant correlation. 

\begin{figure}
    \epsscale{1.1}
    \plotone{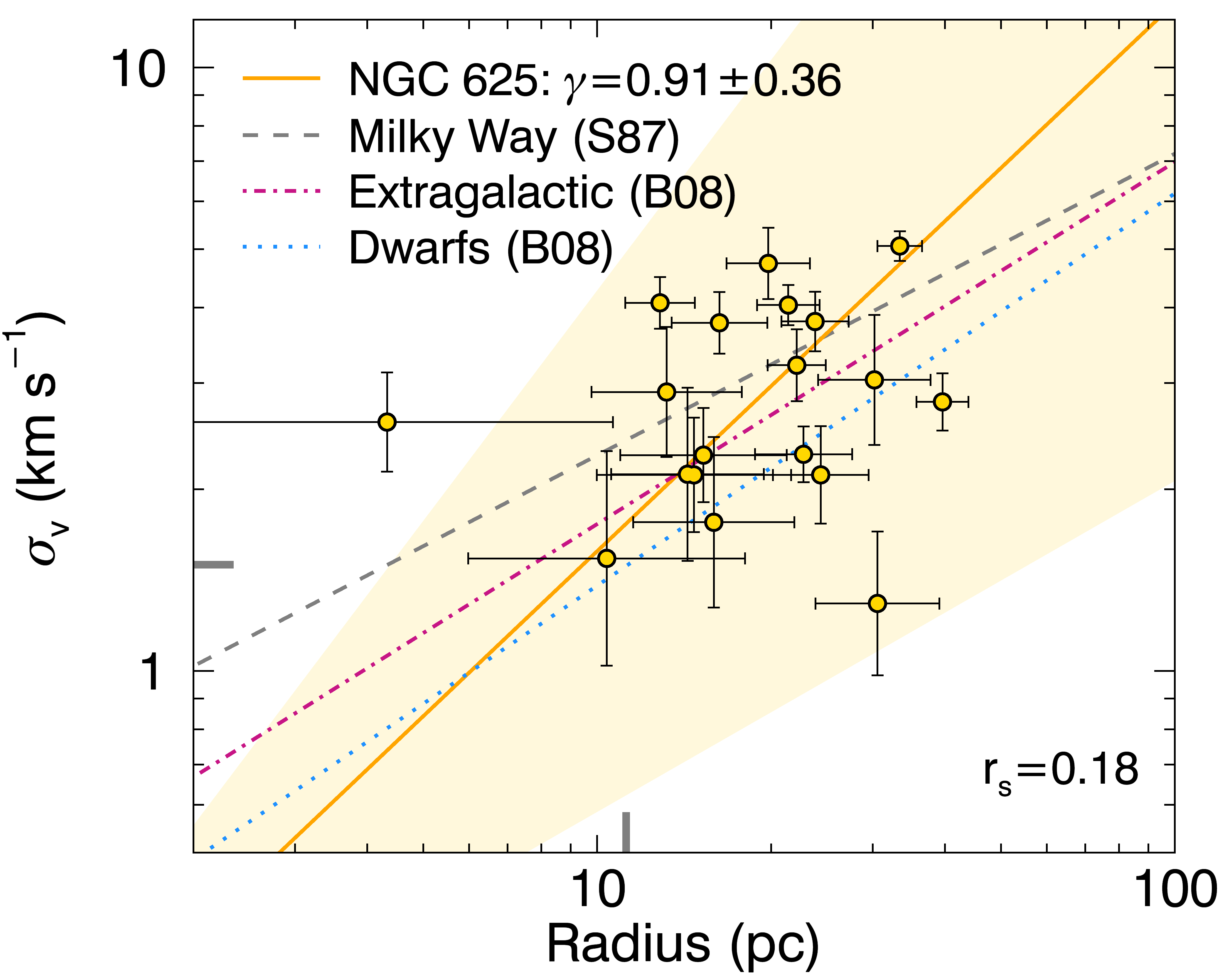}
    \caption{The size-linewidth relation for GMCs in NGC 625.  The solid gold line is the fitted relation using the BCES method, with the shaded region indicating the uncertainty in the slope.  The gray hatch marks indicate the estimated sensitivity limits of the observations.  The Spearman's rank coefficient is shown in the lower right hand corner.  The gray dashed line shows the Milky Way fit: $\sigma_v(\kms)=0.72R(\pc)^{0.5}$ \citepalias{Solomon_1987}.  The pink line shows the fit for nearby galaxies: $\sigma_v(\kms)=0.44R(\pc)^{0.6}$ \citepalias{Bolatto_2008}.  The blue dotted line shows the fit for nearby dwarf galaxies only: $\sigma_v(\kms)=0.31R(\pc)^{0.65}$ \citepalias{Bolatto_2008}.}
    \label{fig:size_linewidth}
\end{figure}

\subsubsection{Size-linewidth relation}\label{sec:sizelw}
Figure \ref{fig:size_linewidth} shows the plot of linewidth versus size for NGC 625 molecular clouds.  Within the $1\sigma$ uncertainty of the fit, the distribution of clouds in $\sigma_v$-$R$ space is consistent with observations of both Milky Way and extragalactic cloud populations.  There is a considerable amount of scatter, and the Spearman's coefficient ($r_s=0.18$) indicates that the correlation is poor. Nevertheless, the formal fit we derive is $\sigma_v\propto R^{0.91\pm0.36}$.

Most of the NGC 625 clouds have sizes, velocity dispersions, and luminous masses above the resolution and completeness limits (see \S\ref{sec:properties}).  There are two resolved clouds with deconvolved radii below the 11 pc resolution limit (Clouds 2 and 14); and there is only one resolved cloud (19) with a deconvolved velocity dispersion less than the smoothed velocity resolution of $1.5$ \kms.  If we exclude these 3 GMCs and only fit data points with values of $R$ and $\sigma_v$ above these limits, the correlation remains poor $(r_s=0.19)$.  This suggests that clouds below the completeness and resolution limits have a negligible effect on the size-linewidth relation.

\begin{figure}
    \epsscale{1.1}
    \plotone{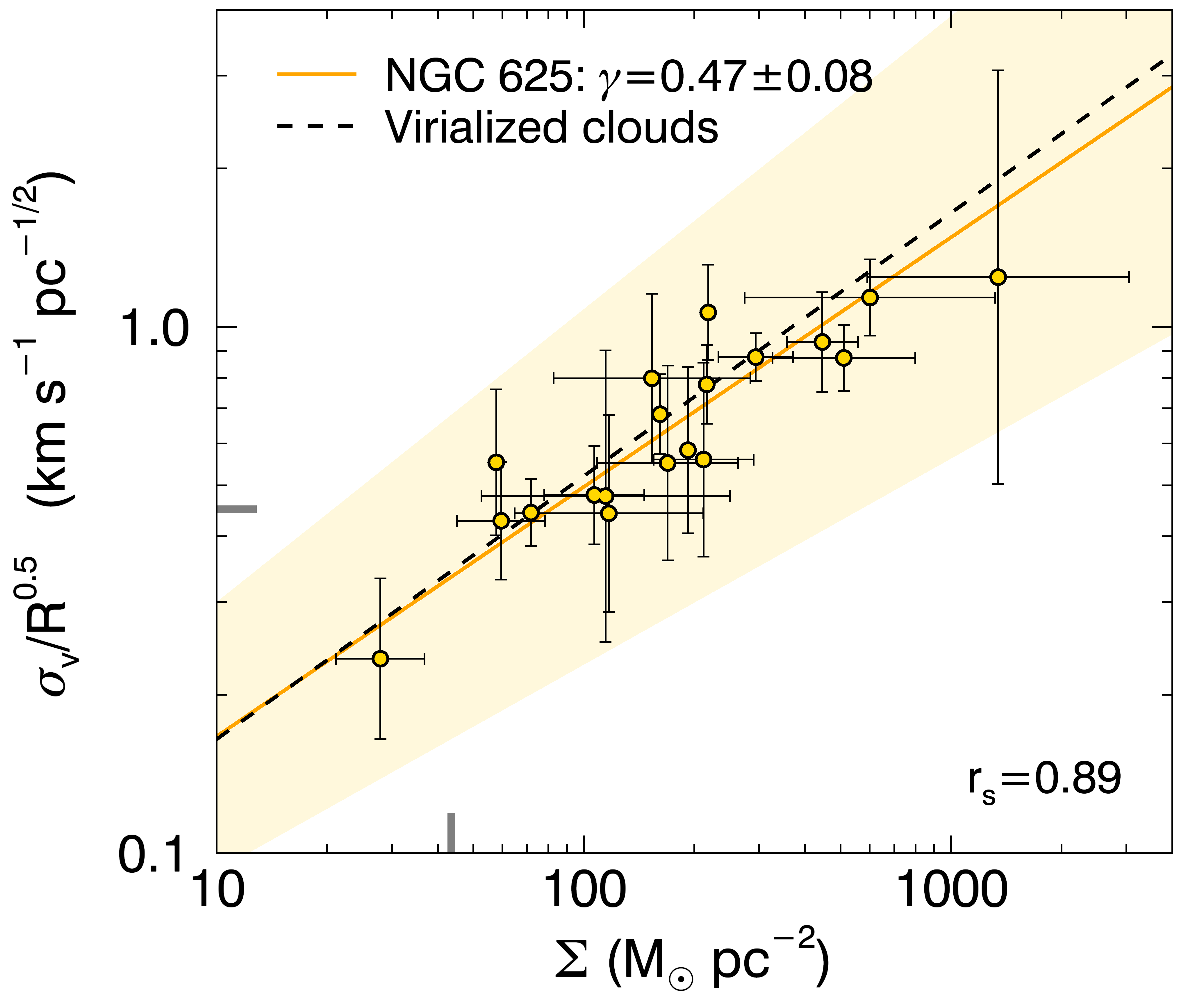}
    \caption{Size-linewidth coefficient versus cloud surface density for NGC\,625.   The solid gold line is the fitted relation using the BCES method, with the shaded region indicating the uncertainty in the slope. The gray hatch marks indicate the estimated sensitivity limits of the observations.  The Spearman's rank coefficient is shown in the lower right hand corner.  The dashed line is the expected relation for virialized clouds.}
    \label{fig:c_sigma}
\end{figure}

\citet{Heyer_2009} found a correlation between cloud mass surface density $\Sigma$ and a size-linewidth coefficient, $C=\sigma_v/R^{1/2}$, in the Boston University-FCRAO Galactic Ring Survey, with a slope expected for gravitationally bound clouds corresponding to $\sigma_v/R^{1/2}=(\pi G/5)^{1/2}\Sigma^{1/2}$.    The $C$-$\Sigma$ plane succinctly encapsulates the Larson's relations.  If all clouds in a system strictly obeyed all three Larson laws, following a size-linewidth relation with the same scaling coefficient and with identical surface densities, this would manifest as a single point located at $\sigma_v/R^{1/2}=(\pi G/5)^{1/2}\Sigma^{1/2}$.  Practically, the presence of observational uncertainties would yield a cluster of points around this location.

Figure \ref{fig:c_sigma} shows that NGC 625 GMCs have a tight correlation between $C$ and $\Sigma$, with a Spearman's coefficient of $0.89$ and a best-fit slope of $0.47\pm0.08$, which is within $1\sigma$ of the prediction for virialized clouds shown by the dashed line, $(\pi G/5)^{1/2}\Sigma^{1/2}=\sigma_v/R^{1/2}$ \citep{Heyer_2009}.  We note that only one cloud has $\Sigma$ less than the 50\% completeness limit of 55 \sunits~(Cloud 19).  And only Clouds 5, 6, 17, and 19 have values of $C<0.45$ \kms~pc$^{-1/2}$, the sensitivity limit we derive for $C$ using our resolution and spectral limits for $\sigma_v$ and $R$ defined above.  If we only consider the 15 clouds above the completness limits, we find that the linear trend is robust, with $r_s=0.86$ and $C\sim\Sigma^{0.37\pm0.1}$. 

This trend implies that the velocity dispersion of clouds does not depend uniquely on the size of the emitting region, but upon surface density as well. The correlation also suggests that the gravitational and kinetic energy of the molecular clouds in NGC 625 are in balance for surface densities spanning an order of magnitude. The $C$-$\Sigma$ correlation emerges naturally when $\Sigma$ is derived from virial masses, since $\mvir/\pi R^2\propto \sigma_v^2/R$.  Yet \citet{Heyer_2009} calculated LTE-derived masses using \co{13} observations to obtain to measure $\Sigma$.  And it is notable that the $C$-$\Sigma$ correlation persists when $\Sigma$ is estimated from \mlum, which assumes a CO-to-\htwo~conversion factor, as in the case with NGC 625 and other extragalactic studies of GMCs \citep[e.g.,][]{Bolatto_2008, Faesi_2018, Imara_2019}.  Considering the appearance of these trends within independent data sets, \citet{Heyer_2009} argued that this evidence is compatible with clouds being in virial equilibrium, and that this is strong proof that the velocity dispersion in molecular clouds depends both upon the area of emission (i.e., their size) and their mass surface density.

\begin{figure}
    \epsscale{1.1}
    \plotone{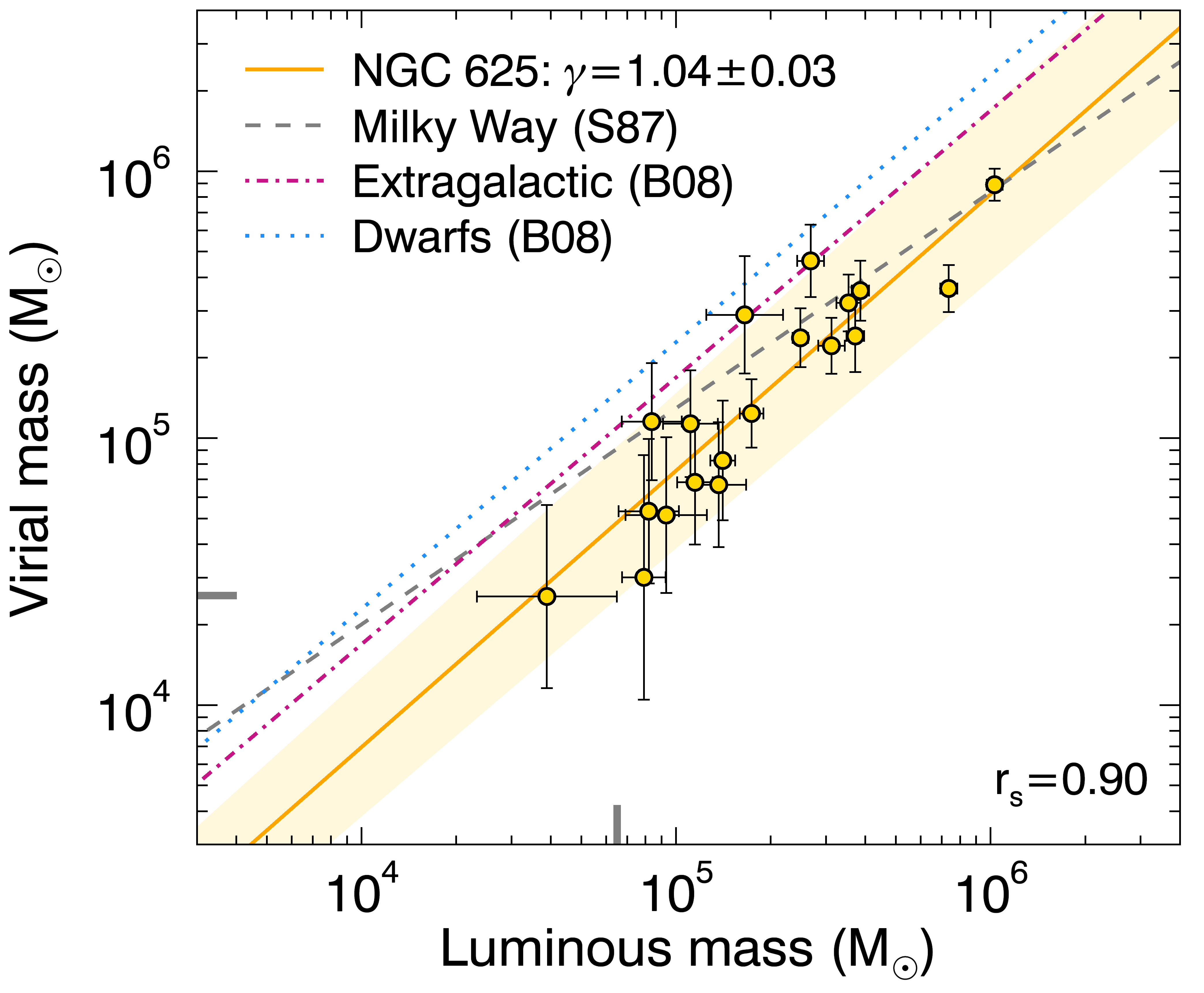}
    \caption{Virial mass versus luminous mass for NGC 625 GMCs.  The solid gold line is the fitted relation using the BCES method, with the shaded region indicating the uncertainty in the slope.  The gray hatch marks indicate the estimated sensitivity limits of the observations. The Spearman's rank coefficient is shown in the lower right hand corner.  The gray dashed line shows the Milky Way fit: $\mvir=11.5\mlum^{0.81}$ \citepalias{Solomon_1987}.  The pink line shows the fit for nearby galaxies: $\mvir=2.2\mlum^{1.0}$ \citepalias{Bolatto_2008}.  The blue dotted line shows the fit for nearby dwarf galaxies only: $\mvir=3.0\mlum^{1.0}$ \citepalias{Bolatto_2008}.}
    \label{fig:virial}
\end{figure}

\subsubsection{Virial relation}\label{sec:virial}
Figure~\ref{fig:virial} shows that there is a strong linear trend between the virial and luminous masses of GMCs in NGC 625, with a correlation coefficient of $r_s=0.90$.  We derive a fit of $\mvir=(0.50\pm0.16)\mlum^{1.04\pm0.03}$, for all 19 resolved clouds.  (Only Cloud 14 has virial and luminous masses below the completeness limits.)

A (nearly) linear correlation has been noticed in several previous studies of Galactic and extragalactic clouds \citep[e.g.,][]{Solomon_1987, Bolatto_2008, Wong_2011, Faesi_2018, Imara_2019}.  Such a trend, as we observe in NGC 625, suggests that clouds are in virial equilibrium.  A high degree of correlation between \mvir~and \mlum~is expected for a moderate range of CO brightness temperatures, $T_{\rm CO}$, since $\mvir\propto R\sigma_v^2$ and if $\mlum\propto R^2 \sigma_v T_{\rm CO}$.  In our case, however, the \mlum, $R$, and $\sigma_v$ we calculate are derived from moments in the CO data cube.  The luminous mass for each cloud is the sum over all pixels over which the cloud is defined, while $R$ and $\sigma_v$ are themselves intensity-weighted moments of the emission distribution (see Equations \ref{eq:radius} and \ref{eq:sigma_v}). Thus, \mlum, $R$, and $\sigma_v$ are correlated but not strictly proportional.

\begin{figure}
    \epsscale{1.1}
    \plotone{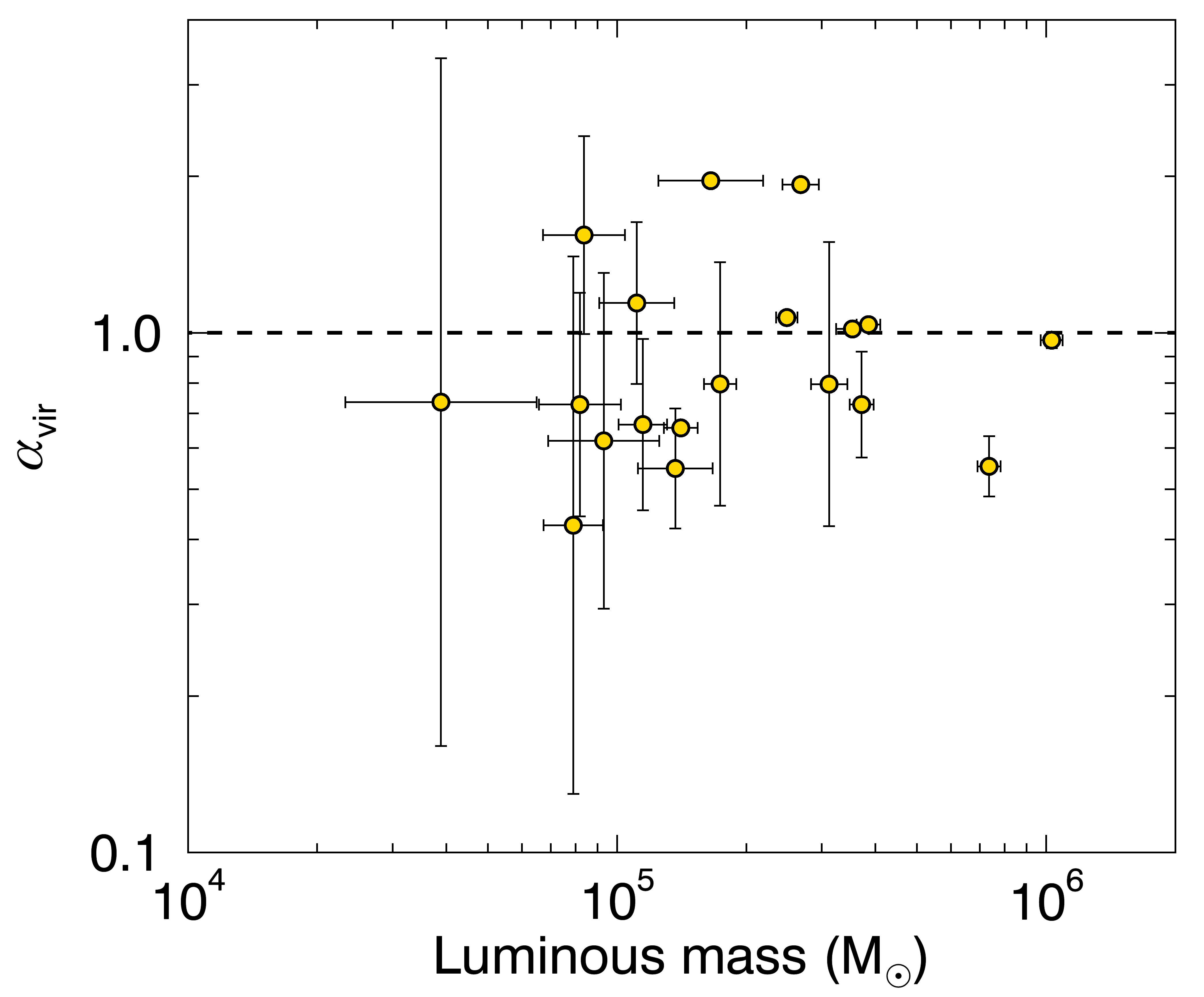}
    \caption{Virial parameter as a function of luminous mass.}
    \label{fig:alpha_vir_mlum}
\end{figure}

In Figure \ref{fig:alpha_vir_mlum} we show the virial parameter, defined 
\begin{equation}\label{eq:virial_param}
    \alpha_{\rm vir}\equiv\frac{5\sigma_v^2 R}{G\mlum}=1.12\frac{\mvir}{\mlum},
\end{equation}
as a function of cloud luminous mass.  Assuming that the CO emission accurately traces cloud mass, the traditional interpretation is that clouds with $\alpha_{\rm vir}\sim1$ are in virial equilibrium, while clouds with $\alpha_{\rm vir}\gg1$ are confined by external pressure \citep{Bertoldi_1992}.  We do not detect evidence suggesting that high-mass clouds tend to be more strongly bound than low-mass clouds, as \citet{Heyer_2001} did for molecular clouds in the outer Galaxy and as \citet{Colombo_2014} found for M51 clouds.

Of course, reliable estimates of the virial parameter and luminous mass depend on a suitable value for the CO-to-\htwo~conversion factor.  At sub-solar metallicity, higher CO-to-\htwo~conversion factors compared to the Milky Way are expected on global scales in galaxies like NGC 625 due to the photodissociation of CO and self-shielding of \htwo~\citep[e.g.,][]{Bolatto_2013}. However, once CO clouds are resolved, it is not clear which conversion factor should apply. Moreover, studies on resolved scales find conversion factors in agreement with the standard Galactic value within a factor of about two \citep[e.g.,][]{Rubio_1993, Indebetouw_2013, Hughes_2013, Schruba_2017, Imara_2019}.  We remind the reader that, insofar as our knowledge of the CO-to-\htwo~conversion factor is incomplete, the \htwo~masses and other conversion-factor-dependent quantities we derive (e.g., surface density) apply to the CO-emitting regions of clouds (see \S\ref{sec:properties}).

\subsubsection{Mass-size relation}

\begin{figure}
    \epsscale{1.1}
    \plotone{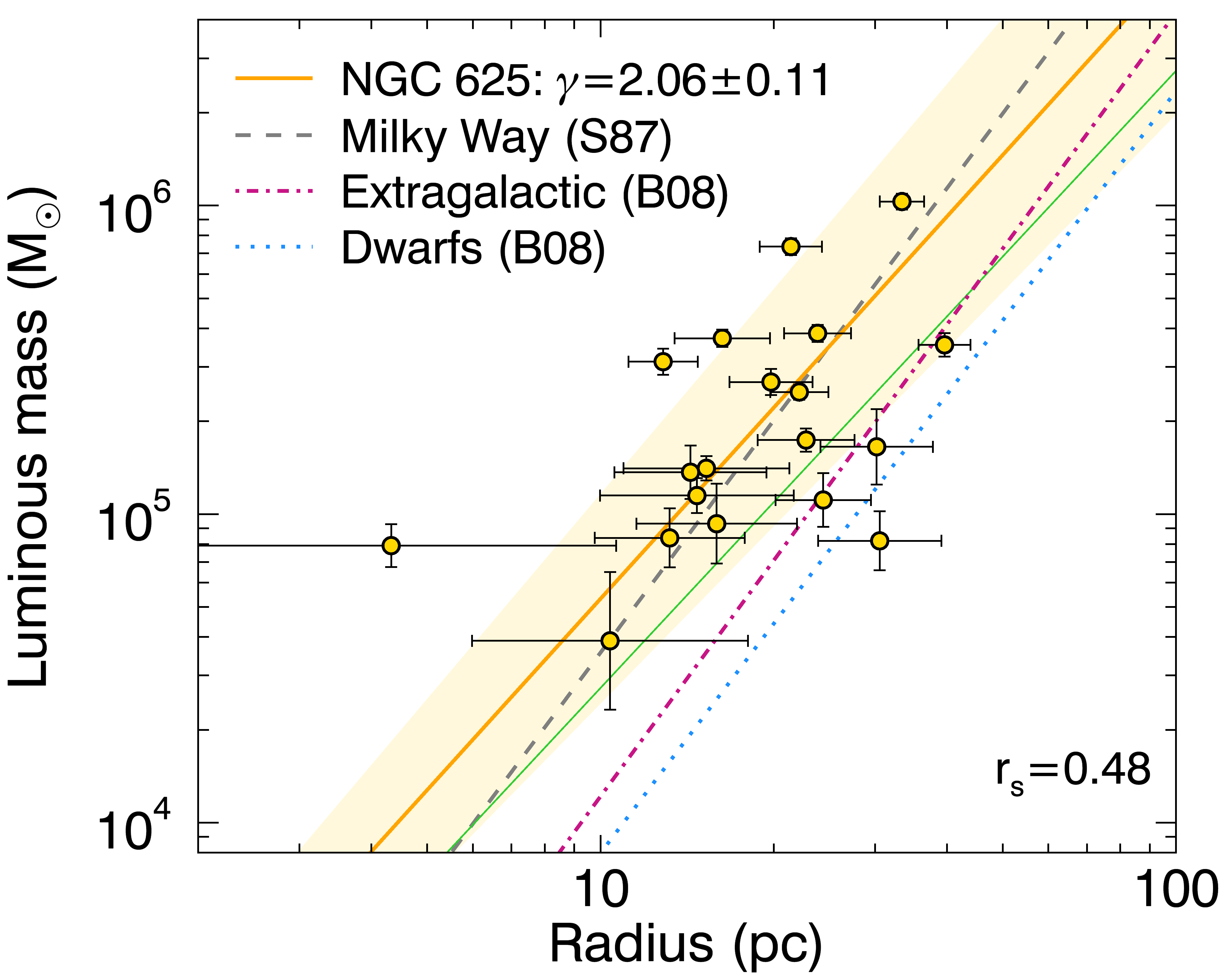}
    \caption{The mass-size relation for GMCs in NGC\,625.  The solid gold line is the fitted relation using the BCES method, with the shaded region indicating the uncertainty in the slope.  The Spearman's rank coefficient is shown in the lower right hand corner.  The green line shows the estimated sensitivity limit for constant mass surface density.  The gray dashed line shows the Milky Way fit: $\mlum=112.5R^{2.5}$ \citep{Solomon_1987}.  The pink line shows the fit for nearby galaxies: $\mlum=35.1R^{2.54}$ \citep{Bolatto_2008}.  The blue dotted line shows the fit for nearby dwarf galaxies only: $\mlum=27.0R^{2.47}$ \citep{Bolatto_2008}.}
    \label{fig:mass_size}
\end{figure}

Figure \ref{fig:mass_size} shows that GMC mass and size are modestly correlated in NGC 625, with $r_s=0.48$.  The vertical offset of the NGC 625 best-fit line relative to the extragalactic trends shows that NGC 625 molecular clouds generally have higher surface densities than clouds in the extragalactic samples compiled by \citetalias{Bolatto_2008}.  On the other hand, the surface densities implied by Figure \ref{fig:mass_size} for NGC 625 clouds are consistent with the typical surface densities observed for Milky Way molecular clouds.    

For Milky Way GMCs, \citet{Larson_1981} identified an inverse correlation between the mean number density, $n$, and size of clouds, $n\sim R^{-1.1}$.  Since $\Sigma\propto M/R^2\propto nR$, this last law implies that molecular clouds have similar surface densities.  The best-fit for NGC 625 clouds shown in Figure \ref{fig:mass_size}, $\mlum = 466 R^{2.06\pm0.11}$, is consistent with constant surface density.   However, previous studies have demonstrated that the slope of $\sim -1$ in the density-size relation measured by \citet{Larson_1981} could be an observational artifact \citep{Kegel_1989, Scalo_1990, Ballesteros-Paredes_2002}.

Two factors may contribute to the apparent trend.  First,  \citet{Kegel_1989} demonstrated that since only clouds with column densities above a certain noise limit are detected, this effectively imposes a constant column density cutoff, which produces a slope of $\sim -1$ in the density-size relation (or $\sim 2$ in the mass-size relation). In Figure \ref{fig:mass_size} we represent the line of constant mass surface density (in green) resulting from our observational sensitivity limit.  Indeed, we do not detect clouds with surface densities below this threshold.  If these clouds exist, it is certainly possible that they would throw off the apparent, albeit weak, trend we now observe.

A second observational artifact may be the limited dynamical range of the observations \citep{Scalo_1990}.  CO observations are sensitive to a restricted range of column densities, and this can manifest as a population of clouds having roughly similar mass surface densities.  The coupling of these two effects and the modest mass-size correlation ($r_s=0.48$) in NGC 625, the $\mlum \propto R^{2.06}$ slope should be interpreted with caution.

\begin{figure}
    \epsscale{1.1}
    \plotone{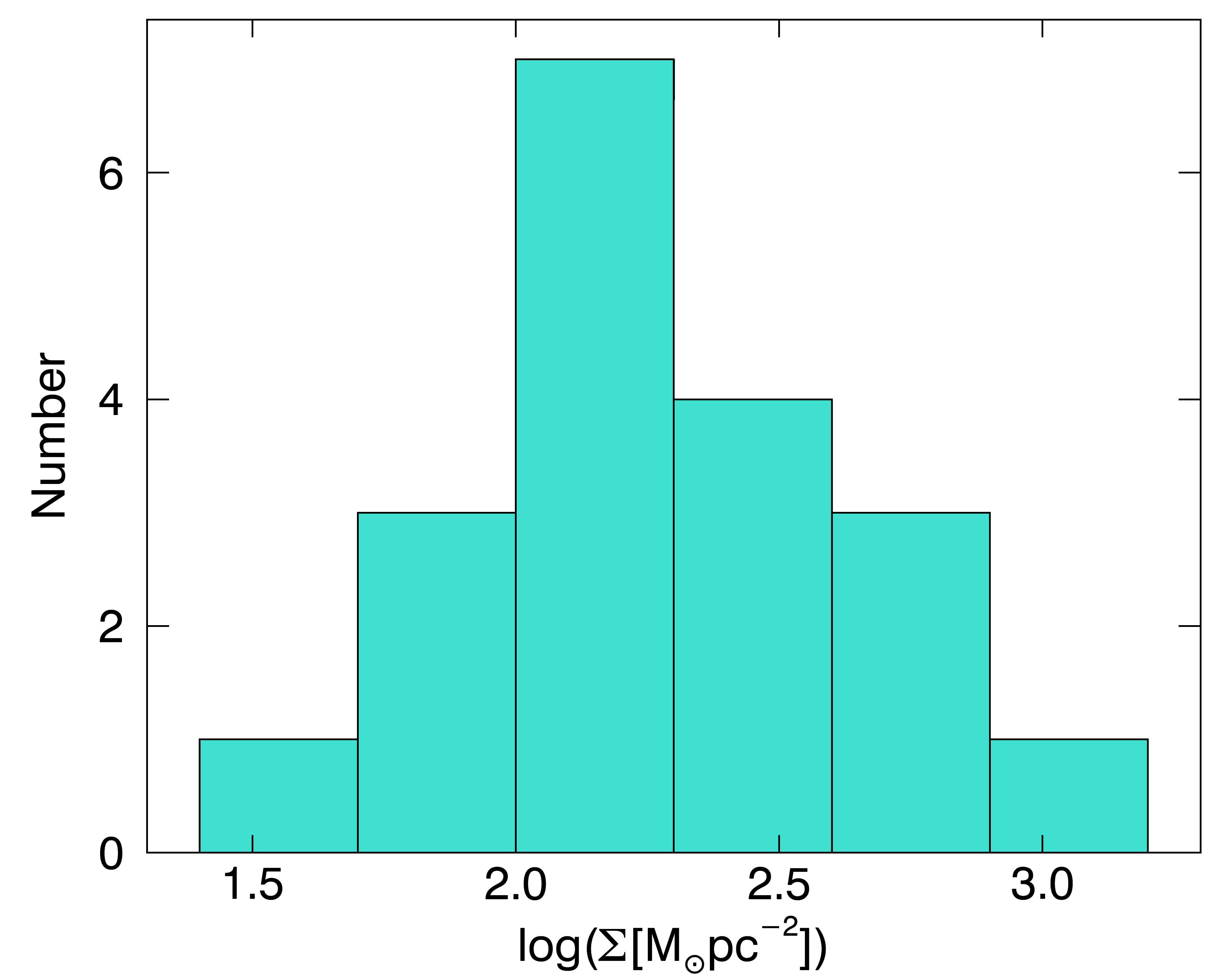}
    \caption{Histogram of GMC mass surface densities in NGC 625.}
    \label{fig:surface_density}
\end{figure}

The considerable scatter in the mass-size distribution is also reflected in the histogram of the GMC mass surface densities presented in Figure \ref{fig:surface_density}. The best-fit to the mass-size relation corresponds to a constant surface density of roughly 148 \sunits. And the median mass surface density of NGC 625 clouds is 169 \sunits, comparable to the $\sim 200$ \sunits~surface density of typical GMCs in the Molecular Ring of the Galaxy \citep{Heyer_2015}.  The range of values, $28$-$1343$ \sunits, span more than an order of magnitude.  Several previous studies have contended that GMCs display a wide range of surface densities in different extragalactic environments \citep[e.g.,][]{Bolatto_2008, Colombo_2014, Utomo_2015, Imara_2019}.  In Sections \ref{sec:distribution} and \ref{sec:sfe}, we explore the variety of environments occupied by clouds having various surface densities.

\begin{figure*}[ht]
    \centering
    \includegraphics[width=6.75in]{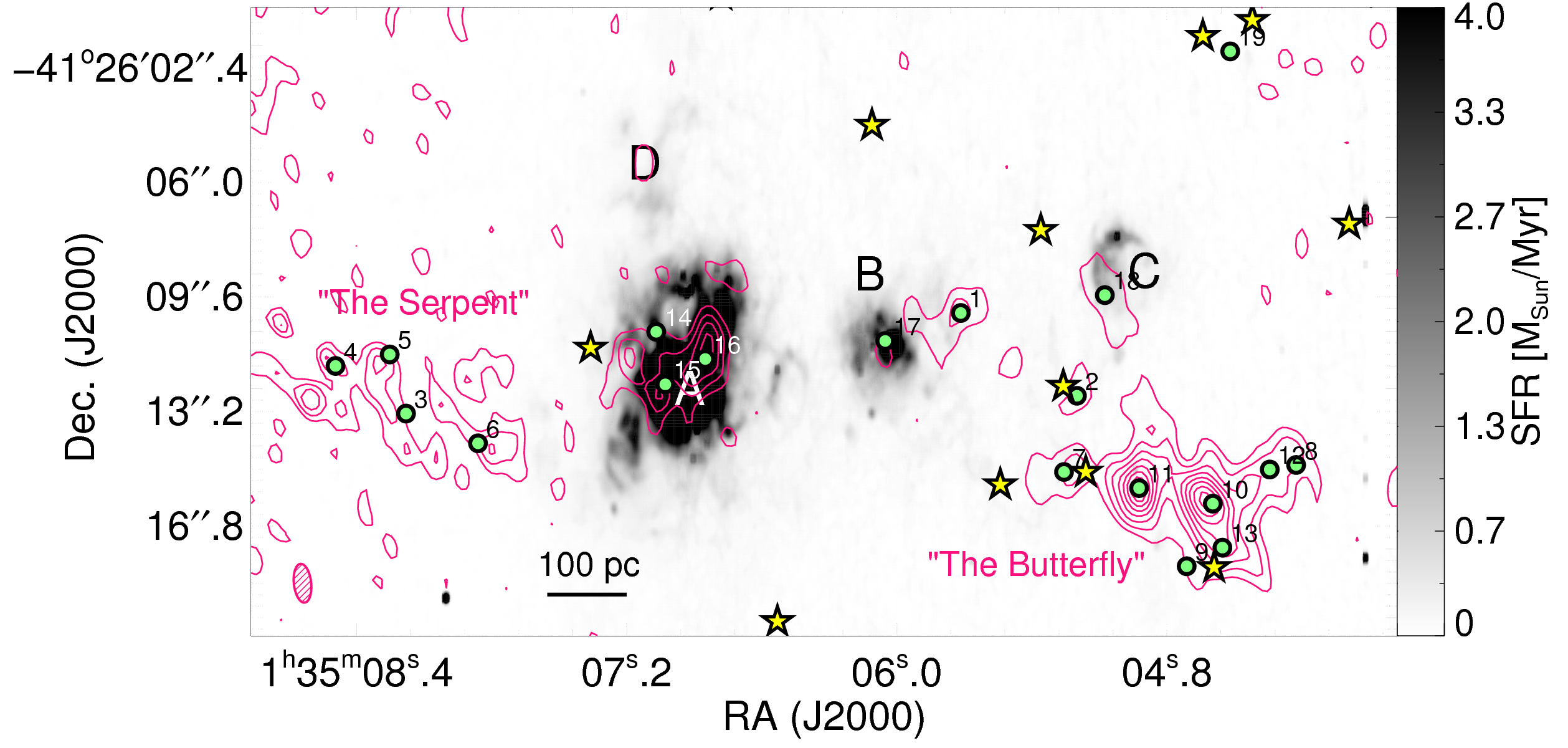}
    \caption{\emph{Hubble Space Telescope} H$\alpha$ image (grayscale) overlaid with the total integrated intensity  ALMA \co{12}(1-0) map (contours) of NGC 625.  The contour range of the CO map is $2$-$14\sigma_{\rm rms}$, where $\sigma_{\rm rms}=7.6$ \counits.  Green circles show the locations of GMCs identified in this paper.  The four brightest \HII~regions are labelled A-D, following \citet{Cannon_2003}.  Additional, low-luminosity regions identified by \citet{Skillman_2003} are shown with yellow stars.  The $1\farcs08\times 1\farcs31$ synthesized beam of the ALMA data is indicated in the lower left.  }
    \label{fig:halpha}
\end{figure*}

\section{Distribution of GMCs and \HII~regions}\label{sec:distribution}
Although Wolf-Rayet signatures are detected in NGC\,625, which might suggest that the galaxy is young \citep[e.g.,][]{Schaerer_1999}, an analysis of its star formation history suggests an SFR that has been steadily decreasing over the past 100 Myr, with a current value of $\simeq0.05$\,$\msun$ yr$^{-1}$ \citep{Cannon_2003,Skillman_2003}. Star formation is mainly concentrated in the dominant H\,{\sc ii} region to the east of the galaxy but some older events are present throughout the galaxy.  There is an age gradient in the stellar population, in the sense that the youngest stars are clustered around the brightest \HII~regions, while the older stars (i.e., asymptotic and red-giant branch stars) are spread more diffusely throughout the galaxy \citep{Cannon_2003}.

Figure \ref{fig:halpha} displays a map of H$\alpha$ emission in NGC 625 observed with the \emph{Hubble Space Telescope} \citep[HST;][]{Cannon_2003} overlaid with our ALMA CO map and the locations of GMCs.  \citet{Cannon_2003} cataloged the four brightest \HII~regions (in terms of their surface brightness, calling them regions A, B, C, and D, which account for $\sim 95\%$ of the total recovered H$\alpha$ flux.  In addition to the four brightest \HII~regions, \citet{Skillman_2003} catalogued several low-luminosity \HII~regions in NGC 625 using ground-based images.  Their locations are indicated by yellow stars in Figure \ref{fig:halpha}.   Our new ALMA observations reveal concentrated peaks in CO emission overlapping spatially with the four regions identified by \citet{Cannon_2003}.  We identify three GMCs each associated with region A, two with region B, one cloud associated with region C, and no clouds associated with region D.  

Six clouds in all, including GMCs 1 and 14-18, are associated with the three brightest \HII~regions identified by \citet{Cannon_2003}.  Five clouds---GMCs 2, 7, 9, 13, and 19---are associated with low-luminosity \HII~regions.   The remaining eight clouds, including GMCs 3-6, 8, and 10-12, are unassociated with any prominent \HII~regions.  We name the complex containing Clouds 3-6, to the east of \HII~region A, the Serpent.

The CO map reveals the presence of concentrations of molecular gas without counterparts in bright \HII~emission, to both the east and west of the galaxy.  The largest and most massive concentration of molecular gas and clouds, located to the southwest of region C, is not associated with a high level of massive star formation activity.  This concentration of molecular gas---which we have named the Butterfly (see Figure \ref{fig:halpha})---is embedded with Clouds 7-12.  Three of these GMCs have the highest luminous masses: Clouds 10, 11, and 13.  These clouds also have the first, third, and fourth highest virial masses, respectively.  

We note that the GMCs in the Butterfly and the Serpent do not depart from the observed Larson relations in NGC 625.  For instance, they are consistent with the observed $\mvir$-$\mlum$ trend.  The GMCs in the Butterfly do have a higher average velocity dispersion, $3.5\pm 0.4$ \kms, than the average $\sigma_v$ of all cataloged clouds ($2.4\pm 0.3$ \kms).   Clouds 7, 10, and 11 have three out of four of the largest one-dimensional velocity dispersions, with values $> 4$ \kms, in the cloud catalog.  The other high-$\sigma_v$ GMC is Cloud 15, associated with \HII~region A. In Table 4, we list the total molecular masses and surface densities of the Serpent and Butterfly.   

\citet{Cannon_2003} found that the youngest stars in NGC 625 are predominantly located near the largest \HII~complexes.  In their Figure 13, they compare the spatial distribution of young, massive main sequence stars (MS; $M_V<-2.5$) with the bright \HII~regions, showing that the most luminous \HII~regions, A and B, are each associated with more than a dozen MS stars, while regions C and D are associated with much fewer.  The map in Figure 13 of \citet{Cannon_2003} also shows the presence of several dozen young stars located to the west of region D and north of regions A and B.  (\citet{Cannon_2003} do not provide the coordinates of the stars.)  

What is the physical structure of molecular clouds that are (un)associated with prominent \HII~regions and young stars in NGC 625?  For the most part, we do not detect any correlations between cloud mass, size, or velocity dispersion and the locations of the brightest \HII~regions.  On the other hand, we find that the highest surface density clouds are clustered in two distinct regions of the galaxy.  The cloud surface density, as defined in Equation (\ref{eq:surf}), ranges from 28 to 1343 \sunits, with a median value of 169 \sunits.  Eight of the highest surface density clouds with $\Sigma>169$ \sunits~are clustered in two regions of the galaxy: in \HII~region A (Clouds 15 and 16) and in the Butterfly (Clouds 7, 8, and 10-13).  Cloud 2 has the highest surface density, with a value of 1343 \sunits.  If we were to define the surface density in terms of the virial mass (as opposed to $\mlum$, as in Equation \ref{eq:surf}), each cloud enumerated above, with the exception of Cloud 8, also has a surface density greater than the median value.

\section{Molecular Depletion Time in NGC 625}\label{sec:sfe}

\subsection{Global scales}\label{sec:global}
NGC 625 has a stellar mass of $3\times 10^8$ \msun~\citep{Madden_2013} and a molecular mass of $5\times 10^6$ \msun~(\S\ref{sec:intensity}), each more than two orders of magnitude lower than those in the Milky Way, which has stellar and molecular masses of $5.2\times 10^{10}\msun$ \citep{Licquia_2015} and $10^9$ \msun~\citep{Heyer_2015}, respectively.  
Moreover, in NGC 625 the \HI~mass \citep[$10^8$ \msun;][]{Cannon_2004} and SFR \citep[0.04-0.05 \msun~yr$^{-1}$][]{Madden_2013, Skillman_2003} are more than an order of magnitude smaller than in the Milky Way, which has an atomic gas mass of $8\times 10^9\msun$ \citep[]{Kalberla_2009} and an SFR of  $1.65~\msun$ yr$^{-1}$ \citep{Licquia_2015}. 

NGC 625 has a gas-to-stellar mass fraction, $f_g\equiv (\mhi + \mhtwo)/\mstar$, roughly twice that of the Milky Way.  For NGC 625, $f_g \approx 0.39$, while in the Milky Way $f_g\approx 0.17$.  This result is consistent with other observations of dwarf galaxies having high gas fractions \citep{deVis_2019}.  Yet the Milky Way may have a greater proportion of its cold neutral gas in the form of molecular gas ($\sim 11\%$) than NGC 625, which has about 5\% of its gas in the form of \htwo. However, we repeat the caveat that the average CO-to-\htwo~conversion factor in NGC 625 may be higher than the Milky Way value we assumed here to estimate the molecular gas mass (see discussion below). The CO-to-\htwo~conversion factor in NGC 625 would need to be only twice the Milky Way value---similar to the LMC \citep[e.g.,][]{Wong_2011}---in order for the galaxy to have a gas fraction on par with the Milky Way.  

The molecular gas depletion time-scale, 
\begin{equation}\label{eq:tdep}
    \tau_{\rm dep}\equiv \frac{\mhtwo}{\rm{SFR}},
\end{equation}
describes the amount of time it would take for a galaxy to use its entire supply of molecular gas at its current rate of star formation, assuming a closed system. In NGC 625, $\tau_{\rm dep}\approx 106$-134 Myr, roughly 5-6 times faster than the  $\sim 600$ Myr depletion time in the Milky Way (using the values for the molecular mass and SFR presented above), and more than an order of magnitude faster than the $\sim 2$ Gyr depletion times measured in other nearby massive spirals \citep{Bigiel_2008}. 

The molecular depletion time we estimate for NGC 625 is also much faster than the $3$ Gyr timescale estimated by \citet{Skillman_2003}, who used the atomic gas mass to estimate $\tau_{\rm dep}$. We reiterate that this depends on the assumption of a Galactic CO-to-\htwo~conversion factor. 
If $\alpha_{\rm CO}$ is larger in NGC 625 and the galaxy has more molecular mass than we infer here, this would imply a longer molecular depletion time.  For the depletion time in NGC 625 to slow down to the $\sim 2$ Gyr measured in massive spirals by \citet{Bigiel_2008}, $\alpha_{\rm CO}$ would have to be at least 10 times higher than the Galactic value.

To better constrain the CO-to-\htwo~conversion factor in NGC 625, we have estimated the conversion factor following \citet{Accurso_2017}, based on its scaling with the [C{\sc{ii}}]/CO ratio. Specifically, we have used Eq. (26) from  \citet{Accurso_2017} using the (PP04) metallicity (8.36, \citealt{deVis_2019}) for this galaxy and its offset from the SF main sequence (using the stellar mass measurement from \citealt{RemyRuyer_2015} and redshift $z=0.00132$ from NED). Based on this, we infer a CO-to-\htwo~conversion factor of $13.23$ \xunits, which is three times higher than a Galactic conversion factor. We furthermore compare the observed [C{\sc{ii}}]/CO ratio ($\log$[C{\sc{ii}}]/CO$ = 4.028$, \citealt{Cormier_2014}) to the predicted value (3.54), using Eq. (16) from \citet{Accurso_2017} based on a galaxy's metallicity and offset from the SF main sequence. As the [C{\sc{ii}}]/CO ratio is known to scale with the CO-to-\htwo~conversion factor, that the observed [C{\sc{ii}}]/CO ratio is larger than the predicted value suggests that the CO-to-\htwo~conversion in NGC 625 could be even higher than three times the Galactic value.

Given the expectation that $\alpha_{\rm CO}$ is expected to increase with decreasing metallicity \citep{Bolatto_2008, Schruba_2012, Cormier_2014}, this is certainly plausible.  Indeed, with a metallicity of $1/3 \zsun$ \citep{Skillman_2003}, assuming that $\alpha_{\rm CO}\propto Z^{-2}$ \citep{Schruba_2012}, the CO-to-\htwo~conversion factor in NGC 625 is $\sim 9$ times the standard Galactic value.

Although the molecular gas fraction and global depletion time in NGC 625 are uncertain, what is unambiguous is that the molecular gas traced by CO and star formation are concentrated in compact regions of the galaxy.  Moreover, \emph{if} the current burst of star formation is short-lived, the galaxy will quickly run out of star-forming fuel if it is not somehow replenished quickly and if its SFR continues at its present rate. In fact, the star formation rate was shown to decline by a factor of $\sim$5 during the last 100\,Myr \citep{Cannon_2003}.

\subsection{Small scales}
\begin{table*}
\centering
\begin{center}
\begin{tabular}{lcccccc}
\multicolumn{7}{c}{Table 4: SFRs and Depletion Times in Select Regions of NGC 625.}\\
\tableline\tableline
Region     & Size of region &  SFR             &   Total \htwo~mass  & Average $\Sigma_{{\rm H}_2}$ & $\tau_{\rm dep}$ & SFE   \\
           & (arcseconds)   & (\msun~yr$^{-1}$)&  ($10^5$ \msun)     &  (\sunits)                   &          & (yr$^{-1}$)        \\  
\tableline
A          & $7\farcs2$ & $2.0\times10^{-2}$   & $8.8$ & 65 & 44 Myr            &  $2.3\times10^{-8}$ \\           
B          & $3\farcs3$ & $1.1\times10^{-2}$   & $6.8$ & 40 & 62 Myr            &  $1.6\times10^{-8}$ \\           
C          & $1\farcs7$ & $7.4\times10^{-4}$   & $1.7$ & 80 & 230 Myr           &  $4.3\times10^{-9}$ \\           
D          & $1\farcs2$ & $6.4\times10^{-4}$   & $0.23$ & 65 & 35 Myr          &  $2.9\times10^{-8}$ \\           
Butterfly  & $9\farcs5\times4\farcs6$ & $\le 5.0\times10^{-4}$ & $25$ & 119 & $\ge 5.0$ Gyr     &  $2.0\times10^{-10}$ \\  
Serpent    & $12\farcs5\times 3\farcs6$ & $\le 5.0\times10^{-4}$ & $7.4$ & 68 & $\ge 1.5$ Gyr   &  $6.6\times10^{-10}$ \\  
\tableline

\end{tabular}
\caption{Column 1: Region name. Column 2: Either the radius or the rectangular dimensions of the region. Column 3: Average SFR.  Column 4: Total \htwo~mass.  Column 4: Average \htwo~surface density.  Column 5: Depletion time. Column 6: Star formation efficiency ($\rm{SFE}=\tau_{\rm dep}^{-1}$).}
\label{table4}
\end{center}
\end{table*}


To study the molecular depletion time scale and the efficiency of star formation in individual H{\sc{ii}} regions, we compare our ALMA \htwo~mass map with the resolved SFR inferred from high-resolution observations with HST (Figure \ref{fig:halpha}).  
Based on this H$\alpha$ map corrected for Galactic and internal dust extinction (\S\ref{sec:hubble}), we infer a map of the total star formation activity in NGC\,625. We note, however, that corrections for internal extinction could only be made in regions where H$\beta$ emission was detected, toward the three brightest \HII~regions A, B, and C (Figure \ref{fig:halpha}).  Outside these regions, the inferred SFR surface density ($\Sigma_{\text{SFR}}$) should be viewed as a lower limit on the true star formation activity (although, the lack of significant MIPS\,24\,$\mu$m emission detected outside of H{\sc{ii}} regions A and B suggest that little embedded star formation is present in those regions).
To convert H$\alpha$ luminosity, $L_{\rm{H}\alpha}$, into a star formation rate, $\Dot{M}_\star$, we use the conversion provided by \citet{Kennicutt_2012}:
\begin{equation}
    \Dot{M}_\star (\msun~\rm{yr}^{-1}) = 10^{-41.27} L_{\rm{H}\alpha} (\rm{erg}~\rm{s}^{-1}).
\end{equation}

We average over the molecular gas mass and SFR within distinct H{\sc{ii}} regions, assuming the sizes of these regions as determined by \citet{Cannon_2003}.  Regions A-D have diameters of $7\farcs2$, $3\farcs3$, $1\farcs7$, and $1\farcs2$, (136, 62, 32, and 23 pc), respectively.
In these H{\sc{ii}} regions, the IMF has been well-sampled to infer a realistic estimate of the SFR averaged over the last 10\,Myr. However, this prescription assumes a constant SFR during the last 10\,Myr, which might not be applicable for NGC\,625. The SFR estimates will therefore be sensitive to the average age of stars in the considered H{\sc{ii}} regions \citep{Leroy_2012}.

In addition, we examine the \htwo~depletion timescale in the Butterfly and the Serpent.  Due to the non-detection of any diffuse H$\alpha$ emission in the HST observations, we assume an upper limit of SFR$\lesssim 5\times10^{-4} \msun$ yr$^{-1}$, inferred from the global H$\alpha$ SFR and the 10$\%$ of diffuse H$\alpha$ emission missed by HST but detected in ground-based observations \citep{Cannon_2003}.  

The depletion times estimated from the total \htwo~masses and SFRs for each region are shown in Table 4, along with the dimensions assumed each region.  We also present the molecular star formation efficiency (SFE) in Table 4, which is simply the inverse of $\tau_{\rm dep}$.  The depletion times associated with the four bright \HII~regions (A-D) range from 35 to 230 Myr, generally consistent with the globally averaged H$_{2}$ depletion time scale of $106-134$ Myr for NGC\,625. For the Butterfly and the Serpent, we infer lower limits on $\tau_{\rm dep}$ of $\ge$5.0\,Gyr and $\ge$1.5\,Gyr, which significantly exceed the typical values inferred for the H{\sc{ii}} regions, and suggests that star formation could still proceed over long timescales within the clouds in these regions. Alternatively, keeping in mind that $\tau_{\rm dep}$ is measured based on the \emph{current} SFR, and there is no apparent star formation in these regions, perhaps star formation has yet to begin in these regions.

The differences between \htwo~depletion timescales in NGC 625 at large and small scales, and the variations we observe between \HII~regions, have also been noted for regions in other galaxies \citep[e.g.,][]{Schruba_2010, Onodera_2010, Liu_2011, Faesi_2014, Kreckel_2018}.  \citet{Schruba_2010} and \citet{Onodera_2010} investigated how the CO-to-H$\alpha$ ratio varies as a function of spatial scale in M33 and found a tight correlation between \htwo~surface density $\Sigma_{\rm{H}_2}$ and SFR per unit area $\Sigma_{\rm SFR}$ on large scales ($\gtrsim 300$ pc), consistent with what has been observed in previous studies \citep{Murgia_2002, Wong_2002, Kennicutt_2007, Bigiel_2008, Wilson_2008}.

On smaller scales, however, the molecular star formation law (i.e., $\Sigma_{\rm SFR}\propto \Sigma_{\rm{H}_2}^n$) breaks down.  \citet{Schruba_2010} and \citet{Onodera_2010} argued that the scale dependence of the \htwo~depletion time results primarily from looking at regions in different evolutionary states.  Similarly, in their study of the molecular star formation efficiency in NGC 628 at spatial scales ranging from 50 to 500 pc, \citet{Kreckel_2018} demonstrated that there is an increase in scatter in the $\Sigma_{\rm SFR}$-$\Sigma_{\rm{H}_2}$ relation as the spatial scale decreases, suggesting that on cloud scales ($\sim 50$ pc), molecular gas and \HII~regions are poorly correlated.  This poor correlation may simply reflect differences in the molecular cloud properties (cloud mass, in particular), across the galaxy, and the different evolutionary states of the star-forming regions.

First, region-to-region differences in the properties of the molecular gas explain some of the variation in molecular depletion times observed on small scales in NGC 625.  As we showed in $\S\ref{sec:distribution}$, the surface densities and masses of GMCs in NGC 625 vary across the galaxy. Indeed, there is more than an order of magnitude range in the total molecular masses of the five regions analyzed here (Table 4), which by definition (Equation \ref{eq:tdep}) contributes to the variation in $\tau_{\rm dep}$.

The different depletion times may also reflect the different ages of GMCs.  In the Milky Way and other galaxies, GMCs display a wide range of star formation activity, which stems from their varying evolutionary stages \citep[e.g.,][]{Kawamura_2009, Chen_2010, Miura_2010, Onodera_2010}.  \citet{Kawamura_2009} categorized GMCs in the LMC into three types, based on their association with massive star formation activity, and interpreted this as an evolutionary sequence.  Those clouds with no signs of massive star formation are the youngest; those with small \HII~regions are currently forming stars; and those associated with both \HII~regions and young stars represent older clouds in the process of being disrupted by the newborn stars.

If this scenario applies in NGC 625, the molecular clouds associated with \HII~regions A-C may be among the oldest in the galaxy.  In their stellar evolution analysis of NGC 625, \citet{Cannon_2003} demonstrated that the youngest stars in the galaxy are associated with the brightest \HII~regions and argued that a previous star formation episode occurred around $50$-$100$ Myr ago. In addition to the lack of bright H$\alpha$ emission, the absence of any recent star formation in the Serpent and the Butterfly has been confirmed by mid-infrared \textit{Spitzer} MIPS24 observations (see Figure \ref{fig:spitzer}, in the Appendix), which suggests that star formation has yet to commence in these parts of the galaxy. In this context, the molecular clouds in the Serpent and the Butterfly may be the youngest and are just beginning to form the next generation of stars.

\section{Summary}\label{sec:summary}
We present new \co{12} ALMA observations of the dwarf starburst galaxy, NGC 625.  These are the highest resolution ($1\arcsec$) observations of molecular gas in this galaxy to date.  We used these data to map the spatial and kinematic structure of the molecular ISM and explore the properties of giant molecular clouds in NGC 625.  We summarize our results:
\begin{enumerate}
    \item 
    The CO line is detected with high signal-to-noise in multiple locations of the galaxy. We reach an rms of 10.3 mJy per 1.5 $\kms$ channel and report a total luminosity of $1.2\times 10^6$ \counits~pc$^2$ and a molecular gas mass of $5.3\times10^6$ \msun, assuming a Galactic conversion factor. There is no ordered velocity structure.
    \item
    CO emission is distributed in discrete clouds. Using the cloud finding algorithm \texttt{CPROPS}, we identify 33 clouds in the map, 19 of which are resolved. NGC\,625 hosts clouds of radii in the range 4-40\,pc, linewidths 1.3-5\,\kms, and luminous masses of 0.04-$1\times10^6$ \msun.  Surface densities range from 28 to $1,343$ \sunits. Our luminous mass and surface density completeness limits derived through false source injection a recovery tests are $6.5\times10^4~\msun$ and $55~\sunits$, respectively.
    \item
    We investigate the Larson's scaling relations and compare cloud properties in NGC 625 to trends observed amongst Milky Way and extragalactic GMCs. We find that the linewidths and sizes of NGC\,625 clouds are, at best, weakly correlated. Measurements of virial and luminous masses suggest that clouds are in virial equilibrium. There is a modest correlation in the mass-size relationship ($\mlum\sim R^2$) implying that clouds have a mass surface densities around 148 \sunits~and that GMCs in this dwarf galaxy have roughly constant surface densities, similar to what we see in massive spiral galaxies.  The $\mlum\sim R^2$ correlation should be interpreted with caution, however, since this could be due to observational effects.  The actual distribution of measured surface densities has a median value of 169 \sunits, comparable to the surface density of typical GMCs in the Molecular Ring of the Milky Way.
    \item
    We compare the location of the clouds with that of bright \HII~regions and stars. About half of the clouds are located towards the brightest \HII~regions, and roughly half are located in two quiescent regions to the southwest (``the Butterfly'') and to the east of the galaxy (``the Serpent'').
    \item
    The global molecular depletion time of the entire galaxy is 106-134\,Myr, assuming a Milky Way CO-to-\htwo~conversion factor $\alpha_{\rm CO}$.  However, if $\alpha_{\rm CO}$ scales with metallicity, the depletion time may be $\sim 9$ times higher.  Star formation near the brightest \HII~regions is rapidly depleting local sources of gas and may be on the decline if the gas is not replenished.  In the Butterfly and Serpent complexes, the depletion times are $>1.5$ Gyr. Clouds in these regions may be younger and could be the fuel for the next generation of stars.
\end{enumerate}

\acknowledgements
This paper makes use of the following ALMA data: ADS/JAO.ALMA\#2015.1.01569.S. ALMA is a partnership of ESO (representing its member states), NSF (USA) and NINS (Japan), together with NRC (Canada) and NSC and ASIAA (Taiwan) and KASI (Republic of Korea), in cooperation with the Republic of Chile. The Joint ALMA Observatory is operated by ESO, AUI/NRAO and NAOJ.

The National Radio Astronomy Observatory is a facility of the National Science Foundation operated under cooperative agreement by Associated Universities, Inc.

This work was supported by the John Harvard Distinguished Science Fellowship at Harvard University.  DC is supported by the European Union's Horizon 2020 research and innovation program under the Marie Sk\l{}odowska-Curie grant agreement No 702622. IDL gratefully acknowledges the support of the Research Foundation Flanders (FWO).

\appendix

\begin{figure*}[ht]
    \centering
    \includegraphics[width=6.25in]{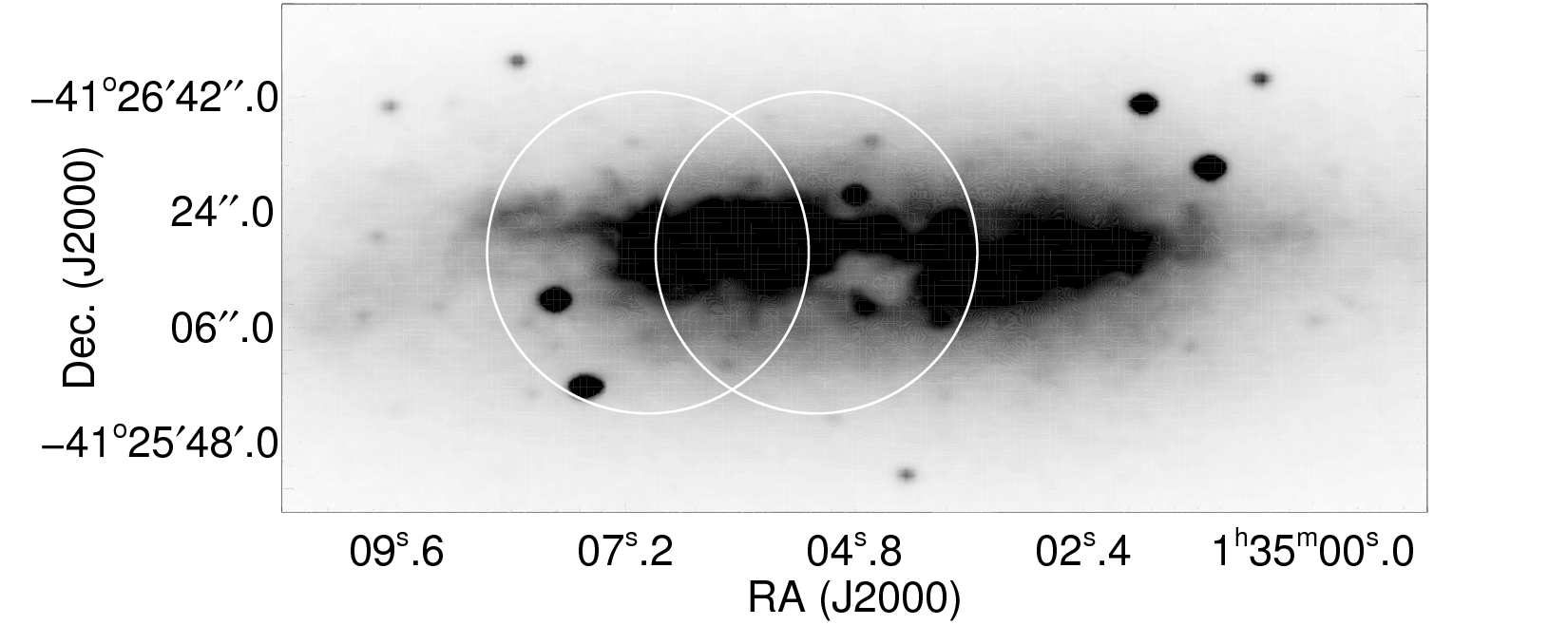}
    \caption{Optical image of NGC 625, overlaid with $50\farcs6 $ ALMA beams (circles).}
    \label{fig:optical}
\end{figure*}

\begin{figure*}[ht]
    \centering
    \includegraphics[width=6.25in]{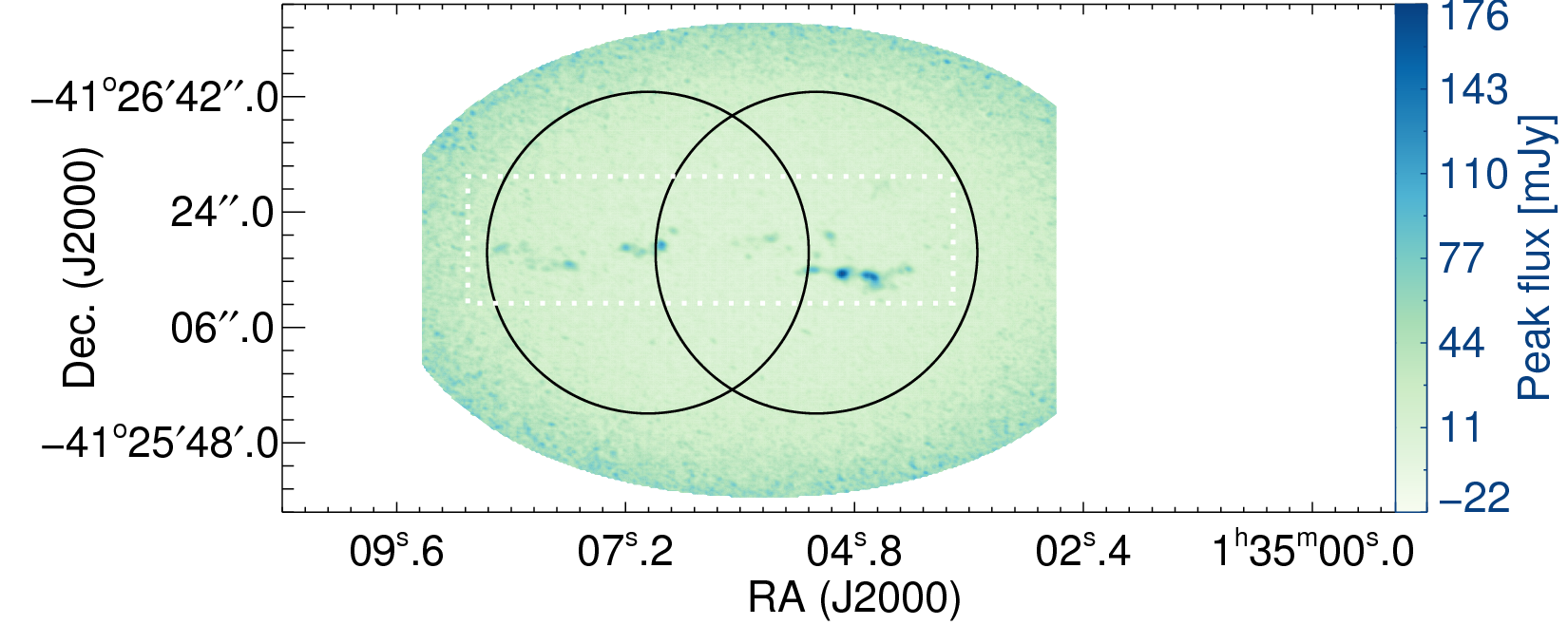}
    \caption{Map of peak flux of our ALMA observations of NGC 625.  The circles represent the $50\farcs6 $ ALMA beams.  The white rectangle indicates the area shown in Figures \ref{fig:map}, \ref{fig:channel}, and \ref{fig:halpha}. }
    \label{fig:peak_flux}
\end{figure*}

\begin{figure*}[ht]
    \centering
    \includegraphics[width=6.25in]{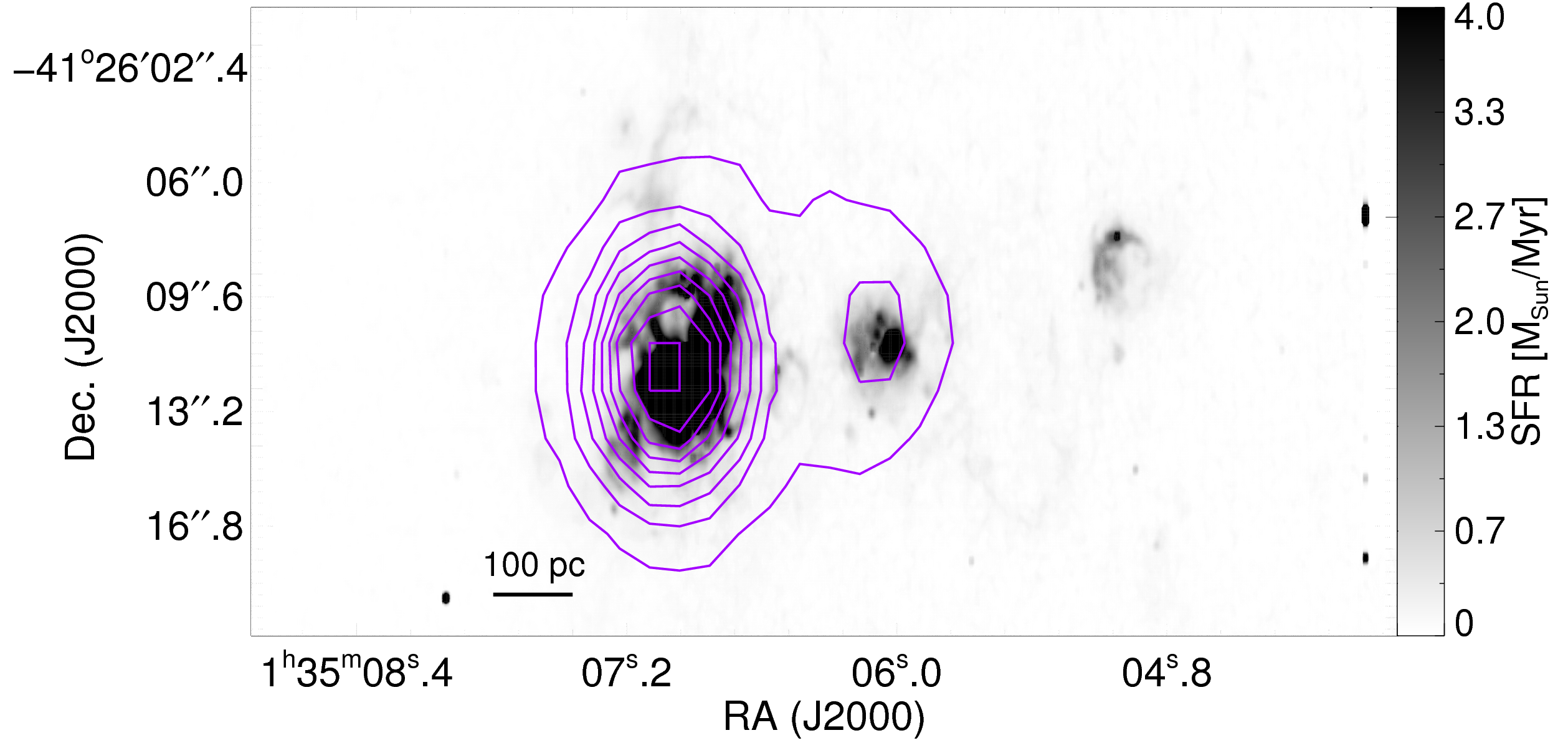}
    \caption{\emph{Hubble} H$\alpha$ image (grayscale) overlaid with  \emph{Spitzer} MIPS $24\micron$ map (contours) of NGC 625.}
    \label{fig:halpha_spitzer}
\end{figure*}

\begin{figure*}[ht]
    \centering
    \includegraphics[width=6.25in]{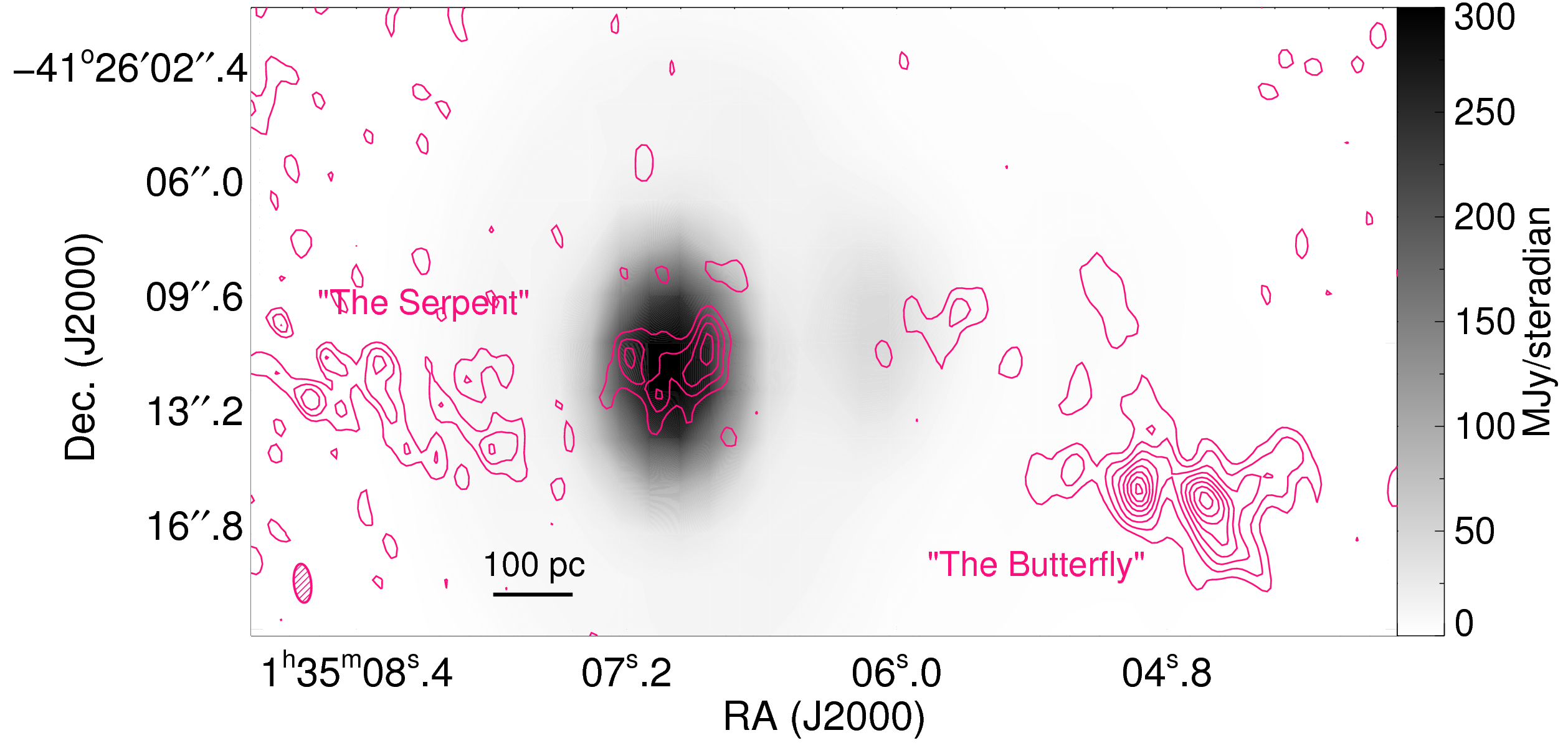}
    \caption{\emph{Spitzer} MIPS $24\micron$ map (grayscale) overlaid with the total integrated intensity  ALMA \co{12}(1-0) map (contours) of NGC 625.  The contour range of the CO map is $2$-$14\sigma_{\rm rms}$, where $\sigma_{\rm rms}=7.6$ \counits.  The $1\farcs08\times 1\farcs31$ synthesized beam of the ALMA data is indicated in the lower left.  }
    \label{fig:spitzer}
\end{figure*}

\bibliography{paper}

\end{document}